\documentclass[%
 reprint,
superscriptaddress,
 amsmath,amssymb,
aps,
pra,
]{revtex4-2}

\usepackage{graphicx}
\usepackage{bm}
\usepackage{amsthm,amsmath,amssymb}
\usepackage{color}
\usepackage{enumitem}
\usepackage{caption}
\usepackage{subcaption}
\usepackage{mathtools}
\usepackage{dcolumn}
\usepackage{hyperref}

\definecolor{Red}{rgb}{1,0,0}

\def\tr{\operatorname{tr}}

\def\poly{\operatorname{poly}}

\begin{document}

\preprint{APS/123-QED}

\title{Quantum transfer component analysis for domain adaptation}

\author{Xi He}
\email{xihe@std.uestc.edu.cn}
\affiliation{Institute of Fundamental and Frontier Sciences, University of Electronic Science and Technology of China}

\author{Chufan Lyu}
\affiliation{Institute of Fundamental and Frontier Sciences, University of Electronic Science and Technology of China}

\author{Min-Hsiu Hsieh}
\email{Min-Hsiu.Hsieh@uts.edu.au}
\affiliation{Center for Quantum Software and Information, Faculty of Engineering and Information Technology, University of Technology Sydney, Australia}

\author{Xiaoting Wang}
\email{xiaoting@uestc.edu.cn}
\affiliation{Institute of Fundamental and Frontier Sciences, University of Electronic Science and Technology of China}

\begin{abstract}
Domain adaptation, a crucial sub-field of transfer learning, aims to utilize known knowledge of one data set to accomplish tasks on another data set. In this paper, we perform one of the most representative domain adaptation algorithms, transfer component analysis (TCA), on quantum devices. Two different quantum implementations of this transfer learning algorithm; namely, the linear-algebra-based quantum TCA algorithm and the variational quantum TCA algorithm, are presented. The algorithmic complexity of the linear-algebra-based quantum TCA algorithm is $O(\poly(\log (n_{s} + n_{t})))$, where $n_{s}$ and $n_{t}$ are input sample size. Compared with the corresponding classical algorithm, the linear-algebra-based quantum TCA can be performed on a universal quantum computer with exponential speedup in the number of given samples.  Finally, the variational quantum TCA algorithm based on a quantum-classical hybrid procedure, that can be implemented on the near term quantum devices, is proposed.
\end{abstract}


\maketitle
\section{Introduction}
\label{sec:introduction}
Quantum computation has the potential to gain speedups in computational complexity compared to the best established classical algorithms~\cite{shor1994algorithms,grover1996fast,harrow2009quantum,aaronson2011computational,farhi2018classification}. When it comes to the subject of machine learning, it has been shown that speedups can also be obtained when the corresponding machine learning algorithms are implemented on a quantum circuit~\cite{lloyd2013quantum,lloyd2014quantum,rebentrost2016quantum}. Specifically, quantum machine learning algorithms can be applied to supervised learning including data fitting~\cite{wiebe2012quantum}, classification~\cite{rebentrost2014quantum}, linear regression~\cite{schuld2016prediction} and unsupervised learning such as clustering~\cite{aimeur2013quantum, wiebe2018quantum}. For the quantum deep learning, quantum restricted Boltzmann machine (qRBM)~\cite{wiebe2014quantum,amin2018quantum}, quantum auto-encoders~\cite{romero2017quantum} and quantum generative adversarial network (QuGAN)~\cite{lloyd2018quantum,dallaire2018quantum} are the representative algorithms. In addition, some algorithms based on the variational quantum eigensolver (VQE)~\cite{peruzzo2014variational} are proposed to deal with machine learning tasks in recent years~\cite{schuld2019quantum,havlivcek2019supervised,higgott2019variational}. 

Transfer learning is an important topic of machine learning and has many applications in computer vision, natural language processing, recommendation system, and hybrid classical-quantum neural networks~\cite{pan2010survey,mari2019transfer}. It is the process of extracting useful information about an unprocessed data set, based on the already-acquired knowledge of a well-studied data set~\cite{pratt1993discriminability}. The key of transfer learning is to identify the connections between the two data sets~\cite{pan2010survey}. Domain adaptation (DA) is a common method to implement transfer learning. It aims to predict the labels of the data from an unlabelled set, based on the known labels of a given data set. Several DA methods have been proposed, namely distribution adaptation, feature selection, and subspace learning. Distribution adaptation maps the distributions of the unlabelled and the labelled data sets, onto a lower-dimensional space, on which the distance of the two induced distributions is minimized~\cite{pan2011domain, long2013transfer, wang2017balanced}. Feature selection is to find the common features of the two data sets~\cite{blitzer2006domain, blitzer2007biographies}, often through implementing certain machine learning algorithms. Subspace learning sequentially transforms the distribution function of the labelled data set subspace so that it gradually approaches to the unlabelled data set subspace~\cite{fernando2013unsupervised, gopalan2011domain, gong2012geodesic}. Transfer component analysis (TCA) is one typical algorithm to implement distribution adaptation. It is based on the assumption that two distributions with similar marginal distributions share similar joint distribution as well, which is often the case for many applications~\cite{pan2011domain}. 

TCA is an effective algorithm in predicting the labels of an unprocessed data set. It transforms the costly kernel learning problem to a procedure of dimensionality reduction. The algorithmic complexity of the original domain adaptation problem is reduced from $O((n_{s}+n_{t})^{6.5})$ to $O(d(n_{s}+n_{t})^{2})$ where $n_{s}$ and $n_{t}$ are the number of the data points in the labelled and unlabelled data sets respectively and $d$ is the dimension of the low-dimensional space we attempt to project to~\cite{pan2011domain}. However, the cost of TCA is prohibitive with the increase of the dimension and number of the data points. 

In this paper, we present quantum versions of the quantum transfer component analysis algorithm (qTCA). In our work, the algorithmic complexity of the linear-algebra-based qTCA is $O(\poly(\log(n_{s}+n_{t})))$. It shows exponential speedup compared with the classical TCA. The linear-algebra-based qTCA can be implemented on a universal quantum computer. In addition, the variational qTCA based on a quantum-classical hybrid procedure is proposed. It can be performed on the near term quantum devices. In the following, we briefly introduce the arrangement of this paper.

In section~\ref{sec:classical_TCA}, we briefly overview the classical TCA algorithm. Then, two implementations of the qTCA algorithm are provided in section~\ref{sec:qTCA}. Having presented all these algorithms, we will perform the quantum support vector machine (qSVM) on the data sets to classify the target labels in section \ref{sec:classification}. The complexity analysis will be discussed in section~\ref{sec:complexity}. Finally, we make a conclusion of all these above.

\section{Classical transfer component analysis}
\label{sec:classical_TCA}
In this section, the basic settings and algorithmic procedures of the classical transfer component analysis algorithm (TCA) will be briefly reviewed.

\begin{figure}[t]
	\includegraphics[width=\columnwidth]{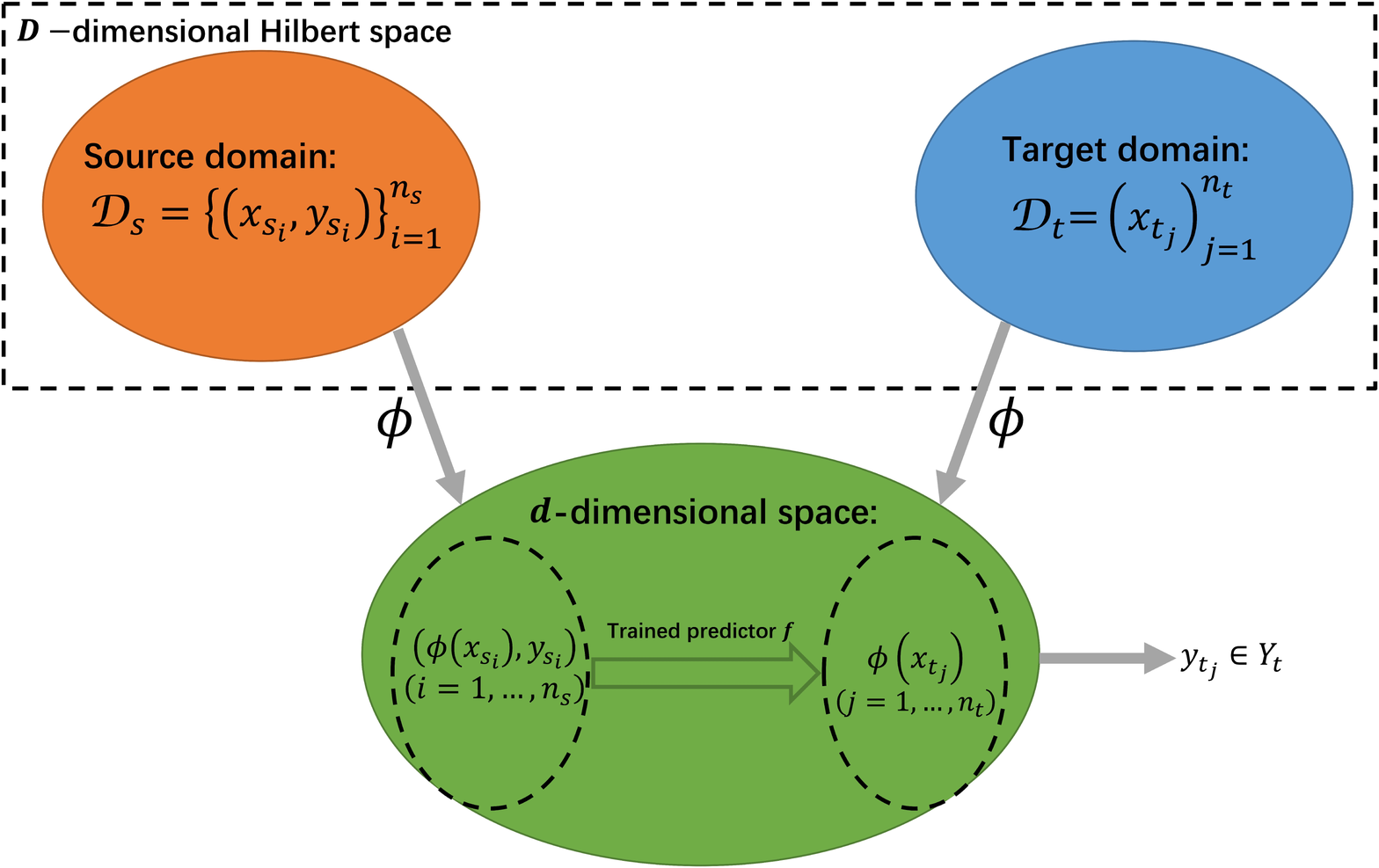}
	\caption{\label{fig:TCA} The schematic diagram of the TCA algorithm.}
\end{figure}

Domain is the main research subject of transfer learning. It mainly contains a data set $\mathcal{D}$ and the probability distribution that generates this data set. Given a labelled source domain data set $\mathcal{D}_s = \{(x_{s_{i}}, y_{s_{i}})\}_{i=1}^{n_{s}}$ and an unlabelled target domain data set $\mathcal{D}_t = \{x_{t_{j}}\}_{j=1}^{n_{t}}$ both in $D$-dimensional space. The data in the source domain and the target domain are independently generated from different distributions. The classical TCA aims to find a feature map $\phi$ to project the original high-dimensional data in the two domains to some $d$-dimensional space ($d \ll D$). Subsequently, a predictor $f$ trained on the source domain data set $\mathcal{D}_{s}$ can be transferred to the target domain data set $\mathcal{D}_{t}$ to predict the target labels $y_t$ $(y_t \in \mathcal{Y}_t)$. The schematic diagram of TCA is depicted in Fig.~\ref{fig:TCA}~\cite{pan2011domain}. 

Generally, the distance between the two distributions in TCA is estimated by the maximum mean discrepancy (MMD)~\cite{gretton2007kernel}
\begin{equation}\label{eq:distance}
    dist(X_s, X_t) = \left \Vert \frac{1}{n_{s}}\sum_{i=1}^{n_{s}}\phi(x_{s_{i}}) - \frac{1}{n_{t}}\sum_{j=1}^{n_{t}}\phi(x_{t_{j}}) \right \Vert,
\end{equation}
where $dist(X_s, X_t)$ represents the distance between $X_s = \{x_{s_{i}}\}_{i=1}^{n_{s}}$ and $X_t = \{x_{t_{j}}\}_{j=1}^{n_{t}}$, and $\phi(\cdot)$ maps $x_{s_{i}}$ and $x_{t_{j}}$ to the reproducing kernel Hilbert space (RKHS)~\cite{smola2007hilbert}.

Subsequently, $dist(X_{s}, X_{t})^{2}$ can be transformed to the matrix form
\begin{align}\label{eq:distance_square}
    dist(X_s, X_t)^{2} &= \left \Vert \frac{1}{n_{s}}\sum_{i=1}^{n_{s}}\phi(x_{s_{i}}) - \frac{1}{n_{t}}\sum_{j=1}^{n_{t}}\phi(x_{t_{j}}) \right \Vert^{2} \notag \\
    &= \tr \left( KL \right)
\end{align}
where the concrete expression of $K$ and $L$ are 
\begin{align}\label{eq:K}
    K &= \begin{bmatrix} \phi(X_{s})^{T} \phi(X_{s})  & \phi(X_{s})^{T} \phi(X_{t}) \\ \phi(X_{t})^{T} \phi(X_{s}) & \phi(X_{t})^{T} \phi(X_{t})
         \end{bmatrix} \nonumber \\
    &= \begin{bmatrix}
    K_{s, s} & K_{s, t} \\
    K_{t, s} & K_{t, t} 
    \end{bmatrix}_{(n_{s}+n_{t}) \times (n_{s}+n_{t})}
\end{align}
and 
\begin{equation}\label{eq:L}
    L = \begin{bmatrix}
    	\frac{1}{n^{2}_{s}} \mathbf{1}_{n_{s}} \mathbf{1}_{n_{s}}^{T} & \frac{-1}{n_{s}n_{t}} \mathbf{1}_{n_{s}} \mathbf{1}_{n_{t}}^{T} \\
    	\frac{-1}{n_{t}n_{s}} \mathbf{1}_{n_{t}} \mathbf{1}_{n_{s}}^{T} & \frac{1}{n^{2}_{t}} \mathbf{1}_{n_{t}} \mathbf{1}_{n_{t}}^{T}
    \end{bmatrix}_{(n_{s}+n_{t}) \times (n_{s}+n_{t})}
\end{equation}
where $\mathbf{1}_{n} = (1, 1,..., 1)^{T}_{n}$ is a $n$-dimensional vector whose elements are all equal to one~\cite{pan2011domain}.

The objective function is defined as  
\begin{equation}\label{eq:objective_function}
    \min_{K \succeq 0} \ \tr(KL) - \lambda\tr(K),
\end{equation}
where $\lambda \geq 0$ is a trade-off parameter~\cite{pan2008transfer}.

With introducing a low-dimensional transformation matrix $W \in \mathbb{R}^{(n_{s}+n_{t}) \times d}$, Eq.~\eqref{eq:objective_function} is transformed to 
\begin{equation}\label{eq:final_goal}
    \begin{split}
        &\min_{W}\ \tr(W^{T}KLKW) + \mu \tr(W^{T}W) \\
        &\quad s.t. \quad W^{T}KHKW = I_d
    \end{split}
\end{equation}
where $\mu > 0$ is a trade-off parameter, and $I_d \in \mathbb{R}^{d \times d}$ is an identity matrix~\cite{pan2011domain}. Specifically, the centering matrix $H = I_{n_{s}+n_{t}} - 1/(n_{s}+n_{t})\mathbf{1}_{n_{s}+n_{t}}\mathbf{1}^{T}_{n_{s}+n_{t}}$. In addition, $HH = H$ and $L^{1/2} = \sqrt{\frac{n_{s}n_{t}}{n_{t}-n_{s}}} L$. 

The optimization problem Eq.~\eqref{eq:final_goal} can be solved with the method of Lagrange multipliers
\begin{equation}\label{eq:lagrange_function}
    \mathcal{L}(W, \Omega) = \tr(W^T(KLK + \mu I)W) - \tr((W^T KHKW - I)\Omega),
\end{equation}
where $\Omega$ is a diagonal matrix whose diagonal elements are respectively corresponding Lagrange multipliers.

Compute the partial derivative of $\mathcal{L}$ with respect to $W$ and set it to zero resulting in 
\begin{equation}\label{eq:derivative}
    (KLK + \mu I)W = KHKW\Omega,
\end{equation}
namely, 
\begin{equation}\label{eq:eigen_equation}
    GW = W\Omega^{-1},
\end{equation}
where $G=(KLK + \mu I)^{-1}KHK$.

Thus, the transformation matrix $W$ can be constructed with the eigenvectors which are corresponding to the $d$ largest eigenvalues of $G$. Ultimately, the final goal of TCA can be reduced to finding the transformation matrix $W$. The low-dimensional data of the source domain and the target domain after dimensionality reduction can be obtained with the transformation matrix $W$, and the predictor $f$ can be subsequently applied on it to predict the labels of the target domain. 

\section{Quantum transfer component analysis}
\label{sec:qTCA}
In this section, the implementations of the quantum transfer component analysis (qTCA) are presented. Firstly, the quantum states are prepared with the quantum linear algebra subroutines in the state preparation section. Then, the qTCA is implemented in two different ways which are the linear-algebra-based qTCA and the variational qTCA respectively.

\subsection{State preparation}
\label{subsec:state_prep}
As introduced in section~\ref{sec:classical_TCA}, what the TCA attempts to do is to find the $d$ largest eigenvalues and the corresponding eigenvectors of $G = (KLK+\mu I)^{-1}KHK$. Let $A = A_{1} + A_{2}$ where $A_{1} = KLK$, $A_{2} = \mu I$ and $B = KHK$. In addition, the data points in the source domain and target domain are combined together to construct $X = \{ x_{s_{1}}, \dots, x_{s_{n_{s}}}, x_{t_{1}}, \dots, x_{t_{n_{t}}} \} = \{ \tilde{x}_{1}, \dots, \tilde{x}_{n_{s}+n_{t}} \} \in \mathbb{R}^{D \times (n_{s}+n_{t})}$. Assume that the data points $\tilde{x}_i$ are stored in the quantum random access memory (qRAM)~\cite{giovannetti2008quantum}. The quantum state $|\tilde{x}_i\rangle = \frac{1}{|\tilde{x}_i|}\sum_{k=1}^{D}(\tilde{x}_{i})_{k}|k\rangle$ can be generated with $q = \log D$ qubits.  

Considering the implementation in practice, the state preparation procedure is initialized with the identity matrix $I$ and the target matrix $G$ can be finally obtained with repeatedly invoking quantum linear algebra subroutines~\cite{harrow2009quantum, lloyd2014quantum, cong2016quantum, duan2017quantum}. It is obvious that $A_{1}$, $A_{2}$ and $B$ are all Hermitian matrices. 

For preparing the state $\rho_{B}$ which is proportional to $B$, $\rho_{0} = \frac{1}{\sqrt{n_{s}+n_{t}}} \sum_{i=1}^{n_{s}+{n_{t}}} | i \rangle \langle i |$ is firstly decomposed in eigenvector basis of the centering matrix $H$ to obtain 
\begin{equation}\label{eq:rho_decomposition}
    \rho_{0} = \sum_{i,j=1}^{n_{s}+n_{t}} \langle u_{i} | \rho_{0} | u_{j} \rangle | u_{i} \rangle \langle u_{j}|,
\end{equation}
where $\{| u_{i} \rangle \}_{i=1}^{n_{s}+n_{t}}$ are the eigenvectors of $H$.

Subsequently, the phase estimation algorithm is applied on $\rho_{0}$ to obtain the state
\begin{equation}\label{eq:operator phase estimation}
	\rho_{1} = \sum_{i,j=1}^{n_{s}+n_{t}} \langle u_{i} | \rho_{0} | u_{j} \rangle | \lambda_{i} \rangle \langle \lambda_{j} | \otimes | u_{i} \rangle \langle  u_{j} |,
\end{equation}
where $\{| \lambda_{i} \rangle \}_{i=1}^{n_{s}+n_{t}}$ represent the eigenvalues of $H$ corresponding to $\{|u_{i} \rangle \}_{i=1}^{n_{s}+n_{t}}$ respectively~\cite{nielsen2002quantum}.

Then, the conditional rotation operation $\textbf{U}_{\textbf{CR}}(H, \lambda)$ is performed on a newly added ancilla qubit to obtain the state
\begin{equation}\label{eq:conditional rotation}
    \rho_{2} = \sum_{i,j=1}^{n_{s}+n_{t}} \langle u_{i} | \rho_{0} | u_{j} \rangle | \lambda_{i} \rangle \langle \lambda_{j} | \otimes | u_{i} \rangle \langle u_{j}|\otimes | \psi_{i} \rangle \langle \psi_{j} |,
\end{equation}
where the ancilla
\begin{equation}\label{eq:psi_1}
	\begin{cases}
	| \psi_{i} \rangle = \sqrt{1 - \gamma_{1}^{2} \lambda_{i}^{2}} | 0 \rangle + \gamma_{1} \lambda_{i} | 1 \rangle; \\
	\langle \psi_{j} | = \sqrt{1 - \gamma_{1^{'}}^{2} {\lambda^{\ast}_{j}}^{2}} \langle 0 | + \gamma_{1^{'}} \lambda_{j}^{\ast} \langle 1 |;
	\end{cases}
\end{equation}
and $\gamma_{1}$, $\gamma_{1^{'}}$ are constants~\cite{harrow2009quantum}.

Finally, we uncompute $|\lambda_{i}\rangle$ to acquire the final state 
\begin{align}\label{eq:uncompute}
    \rho_{3} &= \sum_{i,j=1}^{n_{s}+n_{t}} \langle u_{i} | \rho_{0} | u_{j} \rangle |u_{i} \rangle \langle u_{j} | \otimes | \psi_{i} \rangle \langle \psi_{j} |.
\end{align}

With measuring the ancilla register to be $| 1 \rangle \langle 1 |$, the density operator 
\begin{equation}\label{eq:rho_H}
\rho_{H} = \frac{1}{\sqrt{\sum_{i, j=1}^{n_{s}+n_{t}} \lambda_{i}^{2} {\lambda_{j}^{\ast}}^{2}}} \sum_{i, j=1}^{n_{s}+n_{t}} \lambda_{i} \lambda_{j}^{\ast} \langle u_{i} | \rho_{0} | u_{j} \rangle |u_{i} \rangle \langle u_{j}|
\end{equation}
can be obtained which is proportional to the matrix $HH^{\dagger}$. Because $HH^{\dagger} = H$, $\rho_{H}$ is also proportional to the matrix $H$.

\begin{figure}
	\centering
	\includegraphics[width=\columnwidth]{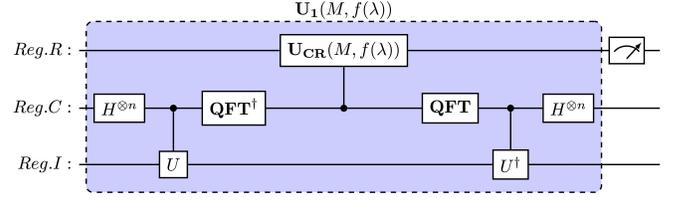}
	\caption{The quantum circuit of $U_{1}(M, f(\lambda))$ where $U = \sum_{\tau=0}^{T-1} | \tau \rangle \langle \tau |^{C} \otimes e^{i M \tau t / T}$ and $n = \log(n_{s} + n_{t})$~\cite{dervovic2018quantum}.}
	\label{fig:U1}
\end{figure}

The whole procedure above can be represented as the following unitary evolution
\begin{align}\label{eq:U_HHL}
	\textbf{U}_{1}(M, f(\lambda)) = &(\textbf{I}^{R} \otimes \textbf{U}^{\dagger}_{\textbf{PE}}(M))(\textbf{U}_{\textbf{CR}}^{RC}(M, f(\lambda)) \otimes \textbf{I}^{I}) \notag \\ 
	&(\textbf{I}^{R} \otimes \textbf{U}_{\textbf{PE}}(M)).
\end{align}
The phase estimation
\begin{align}\label{eq:phase_estimation}
	\textbf{U}_{\textbf{PE}}(M) = &(\textbf{QFT}^{\dagger} \otimes \textbf{I}^{I}) \left( \sum_{\tau=0}^{T-1} | \tau \rangle \langle \tau |^{C} \otimes e^{i M \tau t / T} \right) \notag \\
	&(\textbf{H}^{\otimes n} \otimes \textbf{I}^{I}),
\end{align}
where the matrix $M$ is a specified matrix; $f(\lambda)$ is a specified function of $M$'s eigenvalue $\lambda$ and $\textbf{QFT}^{\dagger}$ stands for the inverse quantum Fourier transform~\cite{duan2017quantum}. And the conditional rotation $\textbf{U}_{\textbf{CR}}(M, f(\lambda))$ is 
\begin{align}\label{eq:U_{CR}}
	&| 0 \rangle \langle 0 |^{R} \otimes | \lambda \rangle \langle \lambda |^{C} \rightarrow | \psi \rangle \langle \psi |^{R} \otimes | \lambda \rangle \langle \lambda |^{C},
\end{align}
where $| \psi \rangle = \sqrt{1 - \gamma^{2} f(\lambda)^{2}} | 0 \rangle + \gamma f(\lambda) | 1 \rangle$; the register $R$ is controlled by the register $C$ and $\gamma$ is a constant~\cite{harrow2009quantum}. The quantum circuit of $\textbf{U}_{1}(M, f(\lambda))$ is depicted in Fig.~\ref{fig:U1}.

\begin{figure*}
	\centering		
	\includegraphics[width=0.9\textwidth]{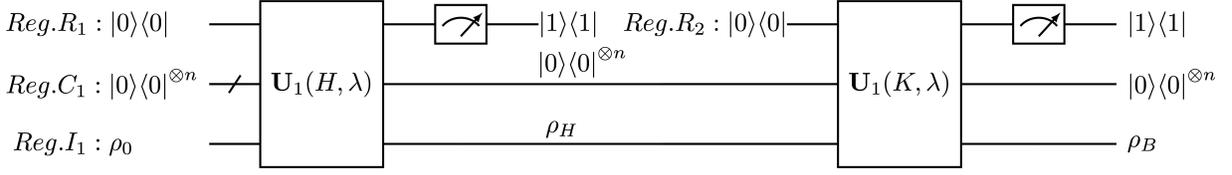}
    \caption{The quantum circuit for preparing $\rho_{B}$.}
    \label{fig:B}
\end{figure*}

Ultimately, the density operator $\rho_{B}$ which is proportional to the matrix $B = KHK$ can be obtained by applying the unitary evolution $\textbf{U}_{1}(K, \lambda)$ on $\rho_{H}$~\cite{cong2016quantum}. The whole procedure of preparing the density operator $\rho_{B}$ is presented as follows
\begin{equation}\label{eq:evolution_B}
	\rho_{0} \xrightarrow{\textbf{U}_{1}(H, \lambda)} \rho_{H} \xrightarrow{\textbf{U}_{1}(K, \lambda)} \rho_{B},
\end{equation}
and the quantum circuit for preparing $\rho_{B}$ is shown in Fig.~\ref{fig:B}.

Similarly, the state $\rho_{A_{1}}$ can be prepared with the unitary evolution as follows:
\begin{equation}\label{eq:evolution_A_{1}}
	\rho_{0} \xrightarrow{\textbf{U}_{1}(L^{1/2}, \lambda)} \rho_{L} \xrightarrow{\textbf{U}_{1}(K, \lambda)} \rho_{A_{1}},
\end{equation}

Because $A = A_{1} + A_{2}$ where $A_{2} = \mu I$. $\rho_{A}$ can be subsequently simulated with utilizing the Suzuki-Trotter expansion~\cite{lloyd2014quantum, lloyd1996universal} as in the following
\begin{equation}\label{eq:Trotter expansion}
	e^{-iAt} = \lim_{k \to \infty}(e^{-i\frac{A_{1}t}{k}}e^{-i\frac{A_{2}t}{k}})^k = e^{-i A_{1} t} e^{-i A_{2} t} + O(t^{2}).
\end{equation}

According to the description in section \ref{sec:classical_TCA}, $G = A^{-1} B$. Thus, the whole process of preparing $\rho_{G}$ is 
\begin{equation}\label{eq:evolution_G}
	\rho_{B} \xrightarrow{\textbf{U}_{1}(A, 1 / \lambda)} \rho_{G},
\end{equation}
where $\rho_{G}$ is proportional to $G$.

In the following, what we need to do is to find the $d$ largest eigenvalues and the corresponding eigenvectors of $\rho_{G}$.  

\subsection{Linear-algebra-based qTCA}
\label{subsec:LABqTCA}
The linear-algebra-based qTCA utilizes the quantum principal component analysis algorithm (qPCA) to estimate the $d$ largest eigenvalues and the corresponding eigenvectors of $G$.

In the first step, a controlled unitary operation $e^{-i \rho_{G} t}$ should be constructed. Ref.~\cite{lloyd2014quantum} proposed an efficient method of Hamiltonian simulation. With the swap operator $S = \sum_{i, j=1}^{n_{s}+n_{t}} | i \rangle \langle j | \otimes | j \rangle \langle i |$, $e^{-i \rho_{G} \Delta t}$ can be applied on the target operator $\sigma$ as follows
\begin{align}\label{eq:rho_evolution}
	\tr_{1}\left( e^{-i S \Delta t} (\rho_{G} \otimes \sigma) e^{i S \Delta t} \right) &= \sigma - i\Delta t[\rho_{G}, \sigma] + O({\Delta t}^2) \notag \\
	&\approx e^{-i \rho_{G} \Delta t} \sigma e^{i \rho_{G} \Delta t}, 
\end{align}   
where $\tr_{1}$ is the partial trace over the first variable and the slice time $\Delta t = t / l$ for some large $l$. The controlled swap operation $CU_{S} = | 0 \rangle \langle 0 | \otimes I + | 1 \rangle \langle 1 | \otimes e^{-i S \Delta t}$. Given $l$ copies of $\rho_{G}$, the controlled unitary operation 
\begin{equation}\label{eq:total_rho_evolution}
	CU_{G} = \sum_{l = 0}^{l} | l \Delta t \rangle \langle l \Delta t | \otimes e^{-i \rho_{G} l \Delta t} \sigma e^{i \rho_{G} l \Delta t}
\end{equation}
can be obtained with repeatedly applying $CU_{S}$ on the states $\sigma \otimes (\rho_{G}^{\otimes l})$~\cite{lloyd2014quantum}. 

With the input state $\rho_{G} = \sum_{i} \lambda_{i} | u_{i} \rangle \langle u_{i} |$ where $u_{i}$ are the eigenvectors of $\rho_{G}$ and $\lambda_{i}$ are the corresponding eigenvalues, it is obvious that $e^{-i \rho_{G} t} | u_{i} \rangle = e^{-i \lambda_{i} t} | u_{i} \rangle$. Subsequently, the quantum phase estimation algorithm can be applied on the input state $\rho_{G}$ resulting in the final state $\sum_{i=1}^{n_{s}+n_{t}} \lambda_{i} | u_{i} \rangle \langle u_{i} | \otimes | \lambda_{i} \rangle \langle \lambda_{i} |$~\cite{nielsen2002quantum}. Ultimately, the transformation matrix $W$ can be obtained with sampling the $d$ dominant eigenvalues and the corresponding eigenvectors from this state. In some cases, $\rho_{G}$ is probably not Hermitian. The quantum phase estimation algorithm can be replaced by some generalized quantum phase estimation algorithms~\cite{wang2010measurement,daskin2014universal}. 

\subsection{Variational quantum transfer component analysis}
\label{subsec:VqTCA}
Although the linear-algebra-based qTCA can be performed on a universal quantum computer, this method requires full coherent evolution and a relatively high quantum circuit depth during the whole process in practice. In order to avoid this required demand, we attempt to design a variational hybrid quantum-classical algorithm inspired by the variational quantum eigensolver (VQE)~\cite{peruzzo2014variational}.

The variational qTCA mainly contains three parts. In the first part, the ansatz state $|\tilde{\psi}(\lambda_{k})\rangle$ is prepared with quantum rotation and entanglement operations where $k = 1, \dots, d$. Afterwards, the quantum data processing will be performed with two parts which are expectation estimation and overlap estimation respectively. At last, the classical section adds all the results of the quantum circuits together and minimizes the cost function resulting in the target eigenstates. 

However, $G$ is probably not Hermitian in some cases. In addition, the goal is to find out $G$'s dominant eigenvalues which is opposite to the purpose of variational quantum computation. Thus, some reformulations should be made. In the first place, we extend $\rho_G$ to
\begin{equation}\label{eq:G_tilde}
	\tilde{G} = \begin{bmatrix}
	0 & -\rho_G \\
	-\rho_{G}^{\dagger} & 0
	\end{bmatrix}_{2(n_{s}+n_{t}) \times 2(n_{s}+n_{t})}
\end{equation}
which is obviously Hermitian. Correspondingly, the original ansatz states should be extended to $| \tilde{\psi} \rangle = | \underbrace{0, \dots, 0}_{n_s + n_t}, \psi \rangle$. In the rest of this section, this algorithm will be alternatively performed with $\tilde{G}$ and $|\tilde{\psi}\rangle$. This variational algorithm will be presented in detail as follows.

\begin{figure}[b]
	\centering
	\includegraphics[width=\columnwidth]{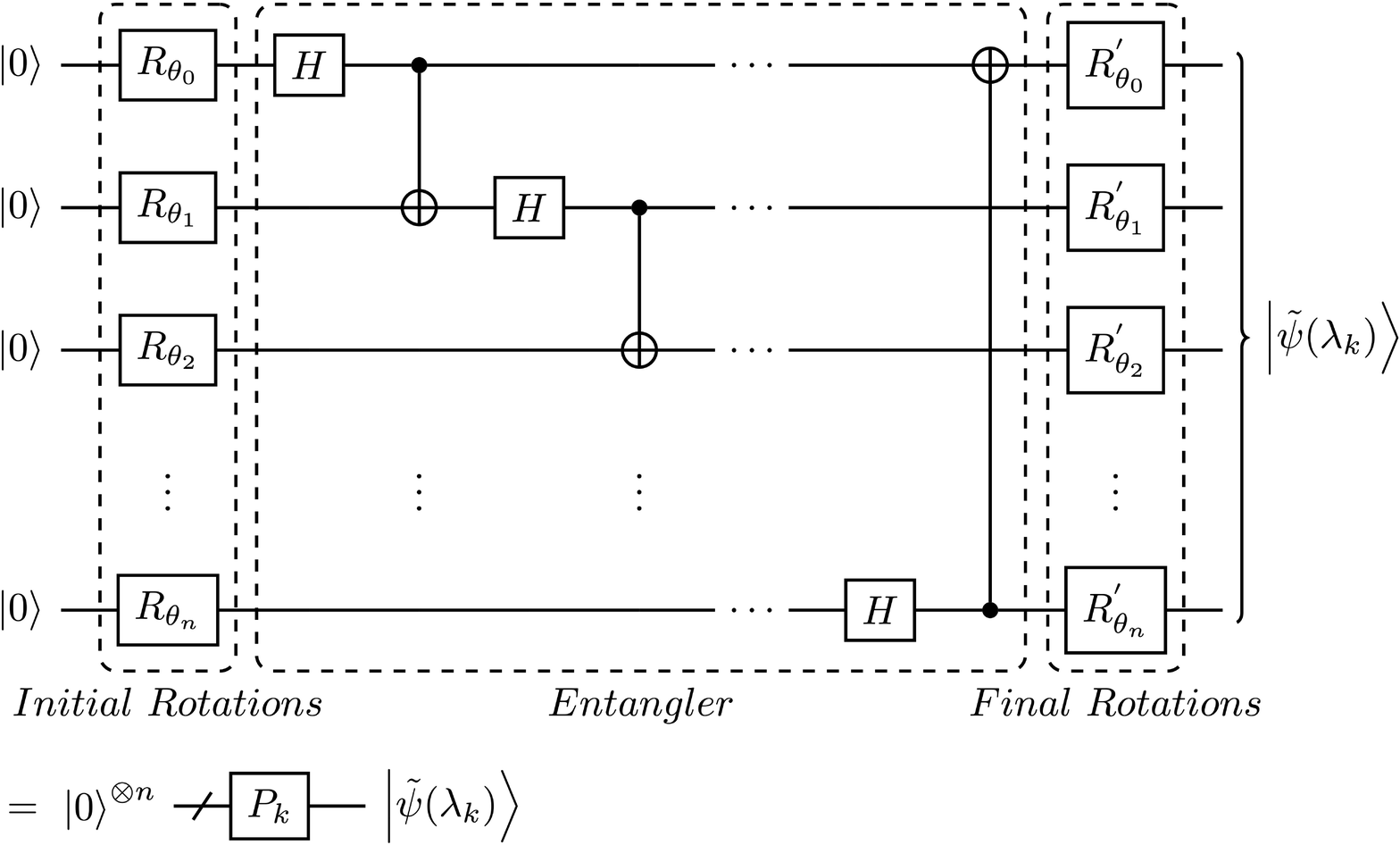} 
\caption{The parameterized quantum circuit of the ansatz state $| \tilde{\psi}(\lambda_{k}) \rangle$ preparation.}
\label{fig:Ansatz}
\end{figure}

(1) The ansatz state $|\tilde{\psi}(\lambda_{k})\rangle$ is constructed with a set of parameterized circuits in practice where $\lambda_{k}$ represents the $k$th eigenvalue of $\tilde{G}$ as shown in Fig.~\ref{fig:Ansatz}. The variational quantum computation algorithm utilizes this state preparation procedure to introduce parameters $\theta$ to the cost function $E(\lambda_{k})$. The preparation circuit of the ansatz state mainly contains three parts which are the initial rotations, the entangler and the final rotations~\cite{du2018expressive}. The rotation operations are a set of parameterized quantum rotation gates. The entangler part can be different in practice and we just present one choice. And the whole preparation circuit is abstracted as a $P_k$ gate.

\begin{figure*}
\centering
\includegraphics[width=0.9\textwidth]{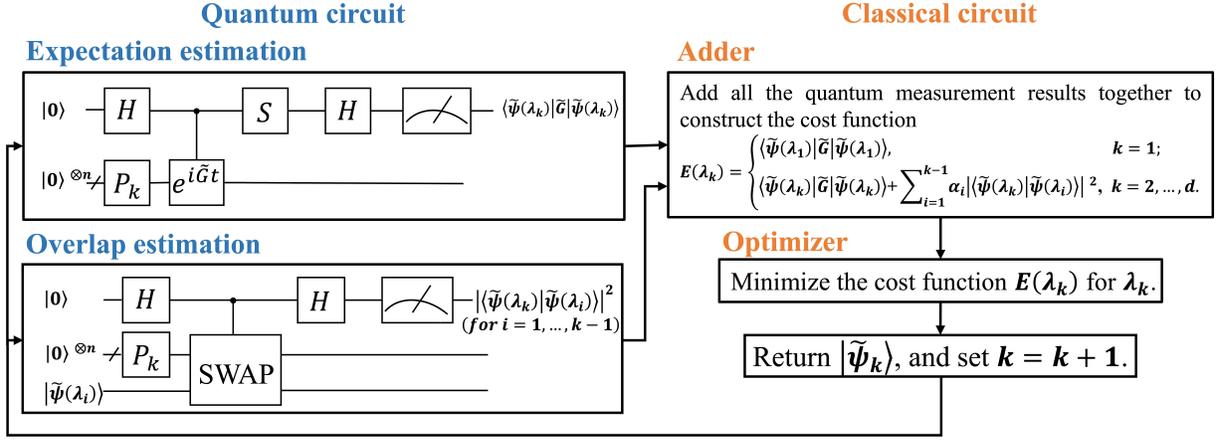}
\caption{The schematic diagram of the variational qTCA.}
\label{fig:VqTCA}
\end{figure*}

(2) Having prepared the ansatz state, we subsequently attempt to compute the eigenstates of $\tilde{G}$ with minimizing the cost function
\begin{widetext}	
\begin{equation}\label{eq:cost_function}
	E(\lambda_{k}) = \begin{cases}
	\langle \tilde{\psi}(\lambda_{1}) | \tilde{G} | \tilde{\psi}(\lambda_{1}) \rangle & k = 1; \\
	\langle \tilde{\psi}(\lambda_{k}) | \tilde{G} | \tilde{\psi}(\lambda_{k}) \rangle + \sum_{i=1}^{k-1} \alpha_{i}|\langle \tilde{\psi}(\lambda_{k}) | \tilde{\psi}(\lambda_{i}) \rangle |^{2} & k = 2, \dots, d,
	\end{cases}
\end{equation}
\end{widetext}
where the expectation value $\langle \tilde{\psi}(\lambda_{k}) | \tilde{G} | \tilde{\psi}(\lambda_{k}) \rangle$ and the overlap term $| \langle \tilde{\psi}(\lambda_{i}) | \tilde{\psi}(\lambda_{k}) \rangle |^2$ are estimated respectively in parallel~\cite{higgott2019variational}. The whole process of solving the $k$th eigenvalue is an iterative process in practice. It should be initiated with computing the minimum of $E_{1} = \langle \tilde{\psi}(\lambda_{1}) | \tilde{G} | \tilde{\psi}(\lambda_{1}) \rangle$. After obtaining $| \tilde{\psi}(\lambda_{1}) \rangle$, we substitute $k = 2$ into Eq.~\eqref{eq:cost_function} resulting in $\lambda_{2}$. And so on, the $k$th eigenvalue can be finally solved out with having obtained the former $k-1$ eigenvalues~\cite{higgott2019variational}. The expectation value term $\langle \tilde{\psi}(\lambda_{k}) | \tilde{G} | \tilde{\psi}(\lambda_{k}) \rangle$ can be estimated with the one step phase estimation quantum circuit as shown in Fig.~\ref{fig:VqTCA}~\cite{roggero2019short}. The estimation of the overlap term $\langle \tilde{\psi}(\lambda_{k}) | \tilde{\psi}(\lambda_{i}) \rangle$ is implemented with the swap test circuit as shown in Fig.~\ref{fig:VqTCA}~\cite{buhrman2001quantum,garcia2013swap,cincio2018learning,zhang2019quantum}. 

(3) The last part is the classical circuit. It contains two parts which are the classical adder and optimizer. The classical adder is introduced on purpose of computing $E(\lambda_{k})$ with summing over all the outputs of the quantum circuits. Then, the cost function $E(\lambda_{k})$ is minimized with a classical optimizer resulting in the value of $\lambda_{k}$. Having obtained $\lambda_{1}, ..., \lambda_{k}$, we can similarly get $\lambda_{k+1}$ by substituting them into the next iteration. Therefore, all the $d$ largest eigenvalues and the corresponding eigenvectors of $G$ are finally obtained with iterating this procedure until $\lambda_{d}$ is determined~\cite{du2018implementable}. The entire process of the variational qTCA is presented in Fig.~\ref{fig:VqTCA}~\cite{he2019quantum}.

\section{Classification}
\label{sec:classification}
As introduced in section~\ref{subsec:LABqTCA} and section~\ref{subsec:VqTCA}, the transformation matrix $W$ can be obtained with the eigenvectors corresponding to the $d$ largest eigenvalues of $G$~\cite{pan2011domain}. Thus, the $d$-dimensional data set $\hat{X} = W^{T} K = \begin{bmatrix} \hat{X}_{s} & \hat{X}_{t} \end{bmatrix}_{d \times (n_{s} + n_{t})}$ where the source domain subspace data set $\hat{X}_{s} \in \mathbb{R}^{d \times n_{s}}$ and the target domain subspace data set $\hat{X}_{t} \in \mathbb{R}^{d \times n_{t}}$. In the following, the quantum support vector machine (qSVM) will be invoked to classify the target labels $y_{t_{j}}$ of the target domain data point $x_{t_{j}}$.

(1) The qSVM classifier model 
\begin{equation}\label{eq:SVM model}
	y(x) = \mathrm{sign}\left( \sum_{i=1}^{n_{s}} \alpha_{i} K_{\xi}(x_{i}, x) + b \right)
\end{equation}
is trained on the low-dimensional source domain subspace data set $\hat{X}_{s}$ and the source domain labels $y_{s}$ to obtain the SVM parameters. 

The SVM parameters $b$ and $\alpha = (\alpha_{1}, \alpha_{2}, \dots, \alpha_{n_{s}})$
\begin{equation}\label{eq:SVM parameters}
	\begin{pmatrix}
		b \\ \alpha
	\end{pmatrix} = \begin{pmatrix}
		0 & \textbf{1}_{n_{s}}^{T} \\
		\textbf{1}_{n_{s}} & K_{\xi} + \xi^{-1} I_{n_{s}}
	\end{pmatrix}^{-1} \begin{pmatrix}
		0 \\ y_{s}
	\end{pmatrix},
\end{equation}
where the kernel matrix $K_{\xi}(x_{i}, x_{j}) = x_{i}^{T} x_{j}$, $y_{s} = (y_{s_{1}}, y_{s_{2}}, \dots, y_{s_{n_{s}}})^{T}$ and $\xi$ is a constant~\cite{rebentrost2014quantum}.

Hence, with the quantum matrix inversion algorithm, the quantum state 
\begin{align}\label{eq:qSVM parameters}
	| b, \alpha \rangle &= \hat{F}^{-1} | y_{s} \rangle = \textbf{U}_{\textbf{1}}(\hat{F}, 1 / \lambda) | y_{s} \rangle \notag \\
	&= \frac{1}{\sqrt{b^{2} + \sum_{i=1}^{n_{s}} \alpha_{i}^{2}}} \left( b|0\rangle + \sum_{i=1}^{n_{s}}\alpha_{i} | k \rangle \right)
\end{align}
can be obtained where $\hat{F} = F / \tr{F}$ with the matrix
\begin{equation}\label{eq:F}
	F = \begin{pmatrix}
		0 & \textbf{1}_{n_{s}}^{T} \\
		\textbf{1}_{n_{s}} & K_{\xi} + \xi^{-1}I_{n_{s}}
	\end{pmatrix}.
\end{equation}
and $| y_{s} \rangle = 1 / |y_{s}|(0, y_{s})^{T}$~\cite{harrow2009quantum}.

(2) The target label $y_{t_{j}}$ can be classified with inputting the low-dimensional target domain subspace data point $\hat{x}_{t_{j}}$ to the qSVM model. The query state
\begin{equation}\label{eq:query state}
	| \psi_{\hat{x}_{t_{j}}} \rangle = \frac{1}{\sqrt{N_{t}}}\left(| 0 \rangle | 0 \rangle + \sum_{i=1}^{n_{s}} | \hat{x}_{t_{j}} | | i \rangle | \hat{x}_{t_{j}} \rangle \right),
\end{equation} 
where $N_{t} = n_{s} | \hat{x}_{t_{j}} |^{2} + 1$. And the quantum state
\begin{equation}\label{eq:training data state}
	| \psi_{t} \rangle = \frac{1}{\sqrt{N_{x}}} \left(b | 0 \rangle | 0 \rangle + \sum_{i=1}^{n_{s}} \alpha_{i} | \hat{x}_{s_{i}} | | i \rangle | \hat{x}_{s_{i}} \rangle \right)
\end{equation}
can be constructed with the training set $\hat{X}_{s}$ where $N_{x} = b^{2} + \sum_{i=1}^{n_{s}} \alpha_{i}^{2} | \hat{x}_{s_{i}} |^{2}$~\cite{rebentrost2014quantum}.

Compute the inner product of $| \psi_{\hat{x}_{t_{j}}} \rangle$ and $| \psi_{t} \rangle$ 
\begin{align}\label{eq:inner product}
	\langle \psi_{t} | \psi_{\hat{x}_{t_{j}}} \rangle &= \frac{1}{\sqrt{N_{t} N_{x}}} \left( b + \sum_{i=1}^{n_{s}} \alpha_{i} | \hat{x}_{s_{i}} | | \hat{x}_{t_{j}} | \langle \hat{x}_{s_{i}} | \hat{x}_{t_{j}} \rangle \right) \notag \\
	&= \frac{1}{\sqrt{N_{t}N_{x}}} \left( b + \sum_{i=1}^{n_{s}} \alpha_{i} K_{\xi}(\hat{x}_{s_{i}}, \hat{x}_{t_{j}}) \right)
\end{align}
with the swap test where $K_{\xi}(\hat{x}_{s_{i}}, \hat{x}_{t_{j}}) = \hat{x}_{s_{i}}^{T} \hat{x}_{t_{j}}$~\cite{buhrman2001quantum}.

The target label
\begin{equation}\label{eq:target label}
	y_{t_{j}} = \mathrm{sign}\left( \sum_{i=1}^{n_{s}} \alpha_{i} K_{\xi}(\hat{x}_{s_{i}}, \hat{x}_{t_{j}}) + b \right)
\end{equation}
can be ultimately classified~\cite{cortes1995support}.

\section{Algorithmic Complexity for Transfer Component Analysis}
\label{sec:complexity}
In this section, the performance of both the classical and quantum TCA algorithms will be evaluated with analyzing and comparing their algorithmic complexity.

In the first place, the algorithmic complexity of the classical TCA algorithm is briefly reviewed. Originally, this domain adaptation problem is considered as a semi-definite programming (SDP) problem which can be solved in $O((n_{s}+n_{t})^{6.5})$ with a kernel learning algorithm~\cite{nesterov1994interior}. The classical TCA algorithm creatively transform the original SDP problem of finding a non-linear map $\phi$ to solving eigenproblems of a specified matrix $G$. The classical TCA algorithm ultimately contributes to reducing the algorithmic complexity to $(d(n_{s}+n_{t})^{2})$~\cite{pan2011domain}.

As to the linear-algebra-based qTCA, let the matrix $H$, $K$ and $L$ can be constructed by visiting a qRAM in time $O(\log (n_{s}+n_{t}))$. Hence, the density operator $\rho_{B}$ can be prepared in time $O((\kappa_{1}^{3}+\kappa_{2}^{4}) \log (n_{s}+n_{t}) / \epsilon^{3})$ where $\kappa_{1}$ and $\kappa_{2}$ are the condition number of $H$ and $K$ respectively and $\epsilon$ is the error parameter~\cite{cong2016quantum}. Similarly, $\rho_{A}$ can be implemented in time $O((\kappa_{3}^{3}+\kappa_{2}^{4}) \log (n_{s}+n_{t}) / \epsilon^{3})$ where $\kappa_{3}$ is the condition number of $L$. Thus, $\rho_{G}$ can be obtained in time $O(\poly(\log(n_{s}+n_{t})), \kappa_{4})$ where $\kappa_{4}$ represents the condition number of $\rho_{A}$~\cite{harrow2009quantum}. With the qPCA, the transformation matrix $W$ can be ultimately obtained in time $O(\log (n_{s}+n_{t}))$. Therefore, the linear-algebra-based qTCA can be implemented in time $O(\poly (\log (n_{s}+n_{t})))$. It is obvious that the linear-algebra-based qTCA can achieve exponential speedup compared with the classical TCA algorithm. In addition, the qSVM algorithm is performed on the low-dimensional data sets to classify the target labels in $O(\log(d n_{s}))$. Compared with the classical SVM, the qSVM can also achieves exponential speedup~\cite{rebentrost2014quantum}. 

\section{Conclusion}
\label{sec:conclusion}
In this paper, we have shown that an important domain adaptation algorithm in transfer learning, the transfer component analysis, can be implemented on the quantum devices. Two quantum versions of the TCA are presented. To be specific, the two designs are respectively based on quantum basic linear algebra subroutines and a variational hybrid quantum-classical process. The linear-algebra-based qTCA can be performed on a universal quantum computer with algorithmic complexity logarithmic in the number of data points. It achieves exponential speedup over classical TCA. In addition, the variational qTCA can be implemented on the near term quantum devices. After the description of the main algorithm, the qSVM algorithm is performed on the low-dimensional data sets to classify the target labels with exponential speedup. Although the linear-algebra-based qTCA requires high quantum circuit depth and the variational qTCA can not provide the quantum speedup in the whole procedure, it is demonstrated that quantum computation can make a contribution to transfer learning. We hope that this work will inspire the design of much more quantum transfer learning algorithms. 

\begin{acknowledgements}
	This work is supported by the National Key R\&D Program of China, Grant No. 2018YFA0306703.
\end{acknowledgements}

\nocite{*}

\bibliography{qTCA}

\begin{thebibliography}{54}%
\makeatletter
\providecommand \@ifxundefined [1]{%
 \@ifx{#1\undefined}
}%
\providecommand \@ifnum [1]{%
 \ifnum #1\expandafter \@firstoftwo
 \else \expandafter \@secondoftwo
 \fi
}%
\providecommand \@ifx [1]{%
 \ifx #1\expandafter \@firstoftwo
 \else \expandafter \@secondoftwo
 \fi
}%
\providecommand \natexlab [1]{#1}%
\providecommand \enquote  [1]{``#1''}%
\providecommand \bibnamefont  [1]{#1}%
\providecommand \bibfnamefont [1]{#1}%
\providecommand \citenamefont [1]{#1}%
\providecommand \href@noop [0]{\@secondoftwo}%
\providecommand \href [0]{\begingroup \@sanitize@url \@href}%
\providecommand \@href[1]{\@@startlink{#1}\@@href}%
\providecommand \@@href[1]{\endgroup#1\@@endlink}%
\providecommand \@sanitize@url [0]{\catcode `\\12\catcode `\$12\catcode
  `\&12\catcode `\#12\catcode `\^12\catcode `\_12\catcode `\%12\relax}%
\providecommand \@@startlink[1]{}%
\providecommand \@@endlink[0]{}%
\providecommand \url  [0]{\begingroup\@sanitize@url \@url }%
\providecommand \@url [1]{\endgroup\@href {#1}{\urlprefix }}%
\providecommand \urlprefix  [0]{URL }%
\providecommand \Eprint [0]{\href }%
\providecommand \doibase [0]{https://doi.org/}%
\providecommand \selectlanguage [0]{\@gobble}%
\providecommand \bibinfo  [0]{\@secondoftwo}%
\providecommand \bibfield  [0]{\@secondoftwo}%
\providecommand \translation [1]{[#1]}%
\providecommand \BibitemOpen [0]{}%
\providecommand \bibitemStop [0]{}%
\providecommand \bibitemNoStop [0]{.\EOS\space}%
\providecommand \EOS [0]{\spacefactor3000\relax}%
\providecommand \BibitemShut  [1]{\csname bibitem#1\endcsname}%
\let\auto@bib@innerbib\@empty
\bibitem [{\citenamefont {Shor}(1994)}]{shor1994algorithms}%
  \BibitemOpen
  \bibfield  {author} {\bibinfo {author} {\bibfnamefont {P.~W.}\ \bibnamefont
  {Shor}},\ }\bibfield  {title} {\bibinfo {title} {Algorithms for quantum
  computation: Discrete logarithms and factoring},\ }in\ \href@noop {} {\emph
  {\bibinfo {booktitle} {Proceedings 35th annual symposium on foundations of
  computer science}}}\ (\bibinfo {organization} {Ieee},\ \bibinfo {year}
  {1994})\ pp.\ \bibinfo {pages} {124--134}\BibitemShut {NoStop}%
\bibitem [{\citenamefont {Grover}(1996)}]{grover1996fast}%
  \BibitemOpen
  \bibfield  {author} {\bibinfo {author} {\bibfnamefont {L.~K.}\ \bibnamefont
  {Grover}},\ }\bibfield  {title} {\bibinfo {title} {A fast quantum mechanical
  algorithm for database search},\ }\href@noop {} {\bibfield  {journal}
  {\bibinfo  {journal} {arXiv preprint quant-ph/9605043}\ } (\bibinfo {year}
  {1996})}\BibitemShut {NoStop}%
\bibitem [{\citenamefont {Harrow}\ \emph {et~al.}(2009)\citenamefont {Harrow},
  \citenamefont {Hassidim},\ and\ \citenamefont {Lloyd}}]{harrow2009quantum}%
  \BibitemOpen
  \bibfield  {author} {\bibinfo {author} {\bibfnamefont {A.~W.}\ \bibnamefont
  {Harrow}}, \bibinfo {author} {\bibfnamefont {A.}~\bibnamefont {Hassidim}},\
  and\ \bibinfo {author} {\bibfnamefont {S.}~\bibnamefont {Lloyd}},\ }\bibfield
   {title} {\bibinfo {title} {Quantum algorithm for linear systems of
  equations},\ }\href@noop {} {\bibfield  {journal} {\bibinfo  {journal}
  {Physical review letters}\ }\textbf {\bibinfo {volume} {103}},\ \bibinfo
  {pages} {150502} (\bibinfo {year} {2009})}\BibitemShut {NoStop}%
\bibitem [{\citenamefont {Aaronson}\ and\ \citenamefont
  {Arkhipov}(2011)}]{aaronson2011computational}%
  \BibitemOpen
  \bibfield  {author} {\bibinfo {author} {\bibfnamefont {S.}~\bibnamefont
  {Aaronson}}\ and\ \bibinfo {author} {\bibfnamefont {A.}~\bibnamefont
  {Arkhipov}},\ }\bibfield  {title} {\bibinfo {title} {The computational
  complexity of linear optics},\ }in\ \href@noop {} {\emph {\bibinfo
  {booktitle} {Proceedings of the forty-third annual ACM symposium on Theory of
  computing}}}\ (\bibinfo {organization} {ACM},\ \bibinfo {year} {2011})\ pp.\
  \bibinfo {pages} {333--342}\BibitemShut {NoStop}%
\bibitem [{\citenamefont {Farhi}\ and\ \citenamefont
  {Neven}(2018)}]{farhi2018classification}%
  \BibitemOpen
  \bibfield  {author} {\bibinfo {author} {\bibfnamefont {E.}~\bibnamefont
  {Farhi}}\ and\ \bibinfo {author} {\bibfnamefont {H.}~\bibnamefont {Neven}},\
  }\bibfield  {title} {\bibinfo {title} {Classification with quantum neural
  networks on near term processors},\ }\href@noop {} {\bibfield  {journal}
  {\bibinfo  {journal} {arXiv preprint arXiv:1802.06002}\ } (\bibinfo {year}
  {2018})}\BibitemShut {NoStop}%
\bibitem [{\citenamefont {Lloyd}\ \emph {et~al.}(2013)\citenamefont {Lloyd},
  \citenamefont {Mohseni},\ and\ \citenamefont
  {Rebentrost}}]{lloyd2013quantum}%
  \BibitemOpen
  \bibfield  {author} {\bibinfo {author} {\bibfnamefont {S.}~\bibnamefont
  {Lloyd}}, \bibinfo {author} {\bibfnamefont {M.}~\bibnamefont {Mohseni}},\
  and\ \bibinfo {author} {\bibfnamefont {P.}~\bibnamefont {Rebentrost}},\
  }\bibfield  {title} {\bibinfo {title} {Quantum algorithms for supervised and
  unsupervised machine learning},\ }\href@noop {} {\bibfield  {journal}
  {\bibinfo  {journal} {arXiv preprint arXiv:1307.0411}\ } (\bibinfo {year}
  {2013})}\BibitemShut {NoStop}%
\bibitem [{\citenamefont {Lloyd}\ \emph {et~al.}(2014)\citenamefont {Lloyd},
  \citenamefont {Mohseni},\ and\ \citenamefont
  {Rebentrost}}]{lloyd2014quantum}%
  \BibitemOpen
  \bibfield  {author} {\bibinfo {author} {\bibfnamefont {S.}~\bibnamefont
  {Lloyd}}, \bibinfo {author} {\bibfnamefont {M.}~\bibnamefont {Mohseni}},\
  and\ \bibinfo {author} {\bibfnamefont {P.}~\bibnamefont {Rebentrost}},\
  }\bibfield  {title} {\bibinfo {title} {Quantum principal component
  analysis},\ }\href@noop {} {\bibfield  {journal} {\bibinfo  {journal} {Nature
  Physics}\ }\textbf {\bibinfo {volume} {10}},\ \bibinfo {pages} {631}
  (\bibinfo {year} {2014})}\BibitemShut {NoStop}%
\bibitem [{\citenamefont {Rebentrost}\ \emph {et~al.}(2016)\citenamefont
  {Rebentrost}, \citenamefont {Steffens},\ and\ \citenamefont
  {Lloyd}}]{rebentrost2016quantum}%
  \BibitemOpen
  \bibfield  {author} {\bibinfo {author} {\bibfnamefont {P.}~\bibnamefont
  {Rebentrost}}, \bibinfo {author} {\bibfnamefont {A.}~\bibnamefont
  {Steffens}},\ and\ \bibinfo {author} {\bibfnamefont {S.}~\bibnamefont
  {Lloyd}},\ }\bibfield  {title} {\bibinfo {title} {Quantum singular value
  decomposition of non-sparse low-rank matrices},\ }\href@noop {} {\bibfield
  {journal} {\bibinfo  {journal} {arXiv preprint arXiv:1607.05404}\ } (\bibinfo
  {year} {2016})}\BibitemShut {NoStop}%
\bibitem [{\citenamefont {Wiebe}\ \emph {et~al.}(2012)\citenamefont {Wiebe},
  \citenamefont {Braun},\ and\ \citenamefont {Lloyd}}]{wiebe2012quantum}%
  \BibitemOpen
  \bibfield  {author} {\bibinfo {author} {\bibfnamefont {N.}~\bibnamefont
  {Wiebe}}, \bibinfo {author} {\bibfnamefont {D.}~\bibnamefont {Braun}},\ and\
  \bibinfo {author} {\bibfnamefont {S.}~\bibnamefont {Lloyd}},\ }\bibfield
  {title} {\bibinfo {title} {Quantum algorithm for data fitting},\ }\href@noop
  {} {\bibfield  {journal} {\bibinfo  {journal} {Physical review letters}\
  }\textbf {\bibinfo {volume} {109}},\ \bibinfo {pages} {050505} (\bibinfo
  {year} {2012})}\BibitemShut {NoStop}%
\bibitem [{\citenamefont {Rebentrost}\ \emph {et~al.}(2014)\citenamefont
  {Rebentrost}, \citenamefont {Mohseni},\ and\ \citenamefont
  {Lloyd}}]{rebentrost2014quantum}%
  \BibitemOpen
  \bibfield  {author} {\bibinfo {author} {\bibfnamefont {P.}~\bibnamefont
  {Rebentrost}}, \bibinfo {author} {\bibfnamefont {M.}~\bibnamefont
  {Mohseni}},\ and\ \bibinfo {author} {\bibfnamefont {S.}~\bibnamefont
  {Lloyd}},\ }\bibfield  {title} {\bibinfo {title} {Quantum support vector
  machine for big data classification},\ }\href@noop {} {\bibfield  {journal}
  {\bibinfo  {journal} {Physical review letters}\ }\textbf {\bibinfo {volume}
  {113}},\ \bibinfo {pages} {130503} (\bibinfo {year} {2014})}\BibitemShut
  {NoStop}%
\bibitem [{\citenamefont {Schuld}\ \emph {et~al.}(2016)\citenamefont {Schuld},
  \citenamefont {Sinayskiy},\ and\ \citenamefont
  {Petruccione}}]{schuld2016prediction}%
  \BibitemOpen
  \bibfield  {author} {\bibinfo {author} {\bibfnamefont {M.}~\bibnamefont
  {Schuld}}, \bibinfo {author} {\bibfnamefont {I.}~\bibnamefont {Sinayskiy}},\
  and\ \bibinfo {author} {\bibfnamefont {F.}~\bibnamefont {Petruccione}},\
  }\bibfield  {title} {\bibinfo {title} {Prediction by linear regression on a
  quantum computer},\ }\href@noop {} {\bibfield  {journal} {\bibinfo  {journal}
  {Physical Review A}\ }\textbf {\bibinfo {volume} {94}},\ \bibinfo {pages}
  {022342} (\bibinfo {year} {2016})}\BibitemShut {NoStop}%
\bibitem [{\citenamefont {A{\"\i}meur}\ \emph {et~al.}(2013)\citenamefont
  {A{\"\i}meur}, \citenamefont {Brassard},\ and\ \citenamefont
  {Gambs}}]{aimeur2013quantum}%
  \BibitemOpen
  \bibfield  {author} {\bibinfo {author} {\bibfnamefont {E.}~\bibnamefont
  {A{\"\i}meur}}, \bibinfo {author} {\bibfnamefont {G.}~\bibnamefont
  {Brassard}},\ and\ \bibinfo {author} {\bibfnamefont {S.}~\bibnamefont
  {Gambs}},\ }\bibfield  {title} {\bibinfo {title} {Quantum speed-up for
  unsupervised learning},\ }\href@noop {} {\bibfield  {journal} {\bibinfo
  {journal} {Machine Learning}\ }\textbf {\bibinfo {volume} {90}},\ \bibinfo
  {pages} {261} (\bibinfo {year} {2013})}\BibitemShut {NoStop}%
\bibitem [{\citenamefont {Wiebe}\ \emph {et~al.}(2018)\citenamefont {Wiebe},
  \citenamefont {Kapoor},\ and\ \citenamefont {Svore}}]{wiebe2018quantum}%
  \BibitemOpen
  \bibfield  {author} {\bibinfo {author} {\bibfnamefont {N.}~\bibnamefont
  {Wiebe}}, \bibinfo {author} {\bibfnamefont {A.}~\bibnamefont {Kapoor}},\ and\
  \bibinfo {author} {\bibfnamefont {K.~M.}\ \bibnamefont {Svore}},\ }\bibfield
  {title} {\bibinfo {title} {Quantum nearest-neighbor algorithms for machine
  learning},\ }\href@noop {} {\bibfield  {journal} {\bibinfo  {journal}
  {Quantum Information and Computation}\ }\textbf {\bibinfo {volume} {15}}
  (\bibinfo {year} {2018})}\BibitemShut {NoStop}%
\bibitem [{\citenamefont {Wiebe}\ \emph {et~al.}(2014)\citenamefont {Wiebe},
  \citenamefont {Kapoor},\ and\ \citenamefont {Svore}}]{wiebe2014quantum}%
  \BibitemOpen
  \bibfield  {author} {\bibinfo {author} {\bibfnamefont {N.}~\bibnamefont
  {Wiebe}}, \bibinfo {author} {\bibfnamefont {A.}~\bibnamefont {Kapoor}},\ and\
  \bibinfo {author} {\bibfnamefont {K.~M.}\ \bibnamefont {Svore}},\ }\bibfield
  {title} {\bibinfo {title} {Quantum deep learning},\ }\href@noop {} {\bibfield
   {journal} {\bibinfo  {journal} {arXiv preprint arXiv:1412.3489}\ } (\bibinfo
  {year} {2014})}\BibitemShut {NoStop}%
\bibitem [{\citenamefont {Amin}\ \emph {et~al.}(2018)\citenamefont {Amin},
  \citenamefont {Andriyash}, \citenamefont {Rolfe}, \citenamefont
  {Kulchytskyy},\ and\ \citenamefont {Melko}}]{amin2018quantum}%
  \BibitemOpen
  \bibfield  {author} {\bibinfo {author} {\bibfnamefont {M.~H.}\ \bibnamefont
  {Amin}}, \bibinfo {author} {\bibfnamefont {E.}~\bibnamefont {Andriyash}},
  \bibinfo {author} {\bibfnamefont {J.}~\bibnamefont {Rolfe}}, \bibinfo
  {author} {\bibfnamefont {B.}~\bibnamefont {Kulchytskyy}},\ and\ \bibinfo
  {author} {\bibfnamefont {R.}~\bibnamefont {Melko}},\ }\bibfield  {title}
  {\bibinfo {title} {Quantum boltzmann machine},\ }\href@noop {} {\bibfield
  {journal} {\bibinfo  {journal} {Physical Review X}\ }\textbf {\bibinfo
  {volume} {8}},\ \bibinfo {pages} {021050} (\bibinfo {year}
  {2018})}\BibitemShut {NoStop}%
\bibitem [{\citenamefont {Romero}\ \emph {et~al.}(2017)\citenamefont {Romero},
  \citenamefont {Olson},\ and\ \citenamefont
  {Aspuru-Guzik}}]{romero2017quantum}%
  \BibitemOpen
  \bibfield  {author} {\bibinfo {author} {\bibfnamefont {J.}~\bibnamefont
  {Romero}}, \bibinfo {author} {\bibfnamefont {J.~P.}\ \bibnamefont {Olson}},\
  and\ \bibinfo {author} {\bibfnamefont {A.}~\bibnamefont {Aspuru-Guzik}},\
  }\bibfield  {title} {\bibinfo {title} {Quantum autoencoders for efficient
  compression of quantum data},\ }\href@noop {} {\bibfield  {journal} {\bibinfo
   {journal} {Quantum Science and Technology}\ }\textbf {\bibinfo {volume}
  {2}},\ \bibinfo {pages} {045001} (\bibinfo {year} {2017})}\BibitemShut
  {NoStop}%
\bibitem [{\citenamefont {Lloyd}\ and\ \citenamefont
  {Weedbrook}(2018)}]{lloyd2018quantum}%
  \BibitemOpen
  \bibfield  {author} {\bibinfo {author} {\bibfnamefont {S.}~\bibnamefont
  {Lloyd}}\ and\ \bibinfo {author} {\bibfnamefont {C.}~\bibnamefont
  {Weedbrook}},\ }\bibfield  {title} {\bibinfo {title} {Quantum generative
  adversarial learning},\ }\href@noop {} {\bibfield  {journal} {\bibinfo
  {journal} {arXiv preprint arXiv:1804.09139}\ } (\bibinfo {year}
  {2018})}\BibitemShut {NoStop}%
\bibitem [{\citenamefont {Dallaire-Demers}\ and\ \citenamefont
  {Killoran}(2018)}]{dallaire2018quantum}%
  \BibitemOpen
  \bibfield  {author} {\bibinfo {author} {\bibfnamefont {P.-L.}\ \bibnamefont
  {Dallaire-Demers}}\ and\ \bibinfo {author} {\bibfnamefont {N.}~\bibnamefont
  {Killoran}},\ }\bibfield  {title} {\bibinfo {title} {Quantum generative
  adversarial networks},\ }\href@noop {} {\bibfield  {journal} {\bibinfo
  {journal} {Physical Review A}\ }\textbf {\bibinfo {volume} {98}},\ \bibinfo
  {pages} {012324} (\bibinfo {year} {2018})}\BibitemShut {NoStop}%
\bibitem [{\citenamefont {Peruzzo}\ \emph {et~al.}(2014)\citenamefont
  {Peruzzo}, \citenamefont {McClean}, \citenamefont {Shadbolt}, \citenamefont
  {Yung}, \citenamefont {Zhou}, \citenamefont {Love}, \citenamefont
  {Aspuru-Guzik},\ and\ \citenamefont {O’brien}}]{peruzzo2014variational}%
  \BibitemOpen
  \bibfield  {author} {\bibinfo {author} {\bibfnamefont {A.}~\bibnamefont
  {Peruzzo}}, \bibinfo {author} {\bibfnamefont {J.}~\bibnamefont {McClean}},
  \bibinfo {author} {\bibfnamefont {P.}~\bibnamefont {Shadbolt}}, \bibinfo
  {author} {\bibfnamefont {M.-H.}\ \bibnamefont {Yung}}, \bibinfo {author}
  {\bibfnamefont {X.-Q.}\ \bibnamefont {Zhou}}, \bibinfo {author}
  {\bibfnamefont {P.~J.}\ \bibnamefont {Love}}, \bibinfo {author}
  {\bibfnamefont {A.}~\bibnamefont {Aspuru-Guzik}},\ and\ \bibinfo {author}
  {\bibfnamefont {J.~L.}\ \bibnamefont {O’brien}},\ }\bibfield  {title}
  {\bibinfo {title} {A variational eigenvalue solver on a photonic quantum
  processor},\ }\href@noop {} {\bibfield  {journal} {\bibinfo  {journal}
  {Nature communications}\ }\textbf {\bibinfo {volume} {5}},\ \bibinfo {pages}
  {4213} (\bibinfo {year} {2014})}\BibitemShut {NoStop}%
\bibitem [{\citenamefont {Schuld}\ and\ \citenamefont
  {Killoran}(2019)}]{schuld2019quantum}%
  \BibitemOpen
  \bibfield  {author} {\bibinfo {author} {\bibfnamefont {M.}~\bibnamefont
  {Schuld}}\ and\ \bibinfo {author} {\bibfnamefont {N.}~\bibnamefont
  {Killoran}},\ }\bibfield  {title} {\bibinfo {title} {Quantum machine learning
  in feature hilbert spaces},\ }\href@noop {} {\bibfield  {journal} {\bibinfo
  {journal} {Physical review letters}\ }\textbf {\bibinfo {volume} {122}},\
  \bibinfo {pages} {040504} (\bibinfo {year} {2019})}\BibitemShut {NoStop}%
\bibitem [{\citenamefont {Havl{\'\i}{\v{c}}ek}\ \emph
  {et~al.}(2019)\citenamefont {Havl{\'\i}{\v{c}}ek}, \citenamefont
  {C{\'o}rcoles}, \citenamefont {Temme}, \citenamefont {Harrow}, \citenamefont
  {Kandala}, \citenamefont {Chow},\ and\ \citenamefont
  {Gambetta}}]{havlivcek2019supervised}%
  \BibitemOpen
  \bibfield  {author} {\bibinfo {author} {\bibfnamefont {V.}~\bibnamefont
  {Havl{\'\i}{\v{c}}ek}}, \bibinfo {author} {\bibfnamefont {A.~D.}\
  \bibnamefont {C{\'o}rcoles}}, \bibinfo {author} {\bibfnamefont
  {K.}~\bibnamefont {Temme}}, \bibinfo {author} {\bibfnamefont {A.~W.}\
  \bibnamefont {Harrow}}, \bibinfo {author} {\bibfnamefont {A.}~\bibnamefont
  {Kandala}}, \bibinfo {author} {\bibfnamefont {J.~M.}\ \bibnamefont {Chow}},\
  and\ \bibinfo {author} {\bibfnamefont {J.~M.}\ \bibnamefont {Gambetta}},\
  }\bibfield  {title} {\bibinfo {title} {Supervised learning with
  quantum-enhanced feature spaces},\ }\href@noop {} {\bibfield  {journal}
  {\bibinfo  {journal} {Nature}\ }\textbf {\bibinfo {volume} {567}},\ \bibinfo
  {pages} {209} (\bibinfo {year} {2019})}\BibitemShut {NoStop}%
\bibitem [{\citenamefont {Higgott}\ \emph {et~al.}(2019)\citenamefont
  {Higgott}, \citenamefont {Wang},\ and\ \citenamefont
  {Brierley}}]{higgott2019variational}%
  \BibitemOpen
  \bibfield  {author} {\bibinfo {author} {\bibfnamefont {O.}~\bibnamefont
  {Higgott}}, \bibinfo {author} {\bibfnamefont {D.}~\bibnamefont {Wang}},\ and\
  \bibinfo {author} {\bibfnamefont {S.}~\bibnamefont {Brierley}},\ }\bibfield
  {title} {\bibinfo {title} {Variational quantum computation of excited
  states},\ }\href@noop {} {\bibfield  {journal} {\bibinfo  {journal}
  {Quantum}\ }\textbf {\bibinfo {volume} {3}},\ \bibinfo {pages} {156}
  (\bibinfo {year} {2019})}\BibitemShut {NoStop}%
\bibitem [{\citenamefont {Pan}\ \emph {et~al.}(2010)\citenamefont {Pan},
  \citenamefont {Yang} \emph {et~al.}}]{pan2010survey}%
  \BibitemOpen
  \bibfield  {author} {\bibinfo {author} {\bibfnamefont {S.~J.}\ \bibnamefont
  {Pan}}, \bibinfo {author} {\bibfnamefont {Q.}~\bibnamefont {Yang}}, \emph
  {et~al.},\ }\bibfield  {title} {\bibinfo {title} {A survey on transfer
  learning},\ }\href@noop {} {\bibfield  {journal} {\bibinfo  {journal} {IEEE
  Transactions on knowledge and data engineering}\ }\textbf {\bibinfo {volume}
  {22}},\ \bibinfo {pages} {1345} (\bibinfo {year} {2010})}\BibitemShut
  {NoStop}%
\bibitem [{\citenamefont {Mari}\ \emph {et~al.}(2019)\citenamefont {Mari},
  \citenamefont {Bromley}, \citenamefont {Izaac}, \citenamefont {Schuld},\ and\
  \citenamefont {Killoran}}]{mari2019transfer}%
  \BibitemOpen
  \bibfield  {author} {\bibinfo {author} {\bibfnamefont {A.}~\bibnamefont
  {Mari}}, \bibinfo {author} {\bibfnamefont {T.~R.}\ \bibnamefont {Bromley}},
  \bibinfo {author} {\bibfnamefont {J.}~\bibnamefont {Izaac}}, \bibinfo
  {author} {\bibfnamefont {M.}~\bibnamefont {Schuld}},\ and\ \bibinfo {author}
  {\bibfnamefont {N.}~\bibnamefont {Killoran}},\ }\bibfield  {title} {\bibinfo
  {title} {Transfer learning in hybrid classical-quantum neural networks},\
  }\href@noop {} {\bibfield  {journal} {\bibinfo  {journal} {arXiv preprint
  arXiv:1912.08278}\ } (\bibinfo {year} {2019})}\BibitemShut {NoStop}%
\bibitem [{\citenamefont {Pratt}(1993)}]{pratt1993discriminability}%
  \BibitemOpen
  \bibfield  {author} {\bibinfo {author} {\bibfnamefont {L.~Y.}\ \bibnamefont
  {Pratt}},\ }\bibfield  {title} {\bibinfo {title} {Discriminability-based
  transfer between neural networks},\ }in\ \href@noop {} {\emph {\bibinfo
  {booktitle} {Advances in neural information processing systems}}}\ (\bibinfo
  {year} {1993})\ pp.\ \bibinfo {pages} {204--211}\BibitemShut {NoStop}%
\bibitem [{\citenamefont {Pan}\ \emph {et~al.}(2011)\citenamefont {Pan},
  \citenamefont {Tsang}, \citenamefont {Kwok},\ and\ \citenamefont
  {Yang}}]{pan2011domain}%
  \BibitemOpen
  \bibfield  {author} {\bibinfo {author} {\bibfnamefont {S.~J.}\ \bibnamefont
  {Pan}}, \bibinfo {author} {\bibfnamefont {I.~W.}\ \bibnamefont {Tsang}},
  \bibinfo {author} {\bibfnamefont {J.~T.}\ \bibnamefont {Kwok}},\ and\
  \bibinfo {author} {\bibfnamefont {Q.}~\bibnamefont {Yang}},\ }\bibfield
  {title} {\bibinfo {title} {Domain adaptation via transfer component
  analysis},\ }\href@noop {} {\bibfield  {journal} {\bibinfo  {journal} {IEEE
  Transactions on Neural Networks}\ }\textbf {\bibinfo {volume} {22}},\
  \bibinfo {pages} {199} (\bibinfo {year} {2011})}\BibitemShut {NoStop}%
\bibitem [{\citenamefont {Long}\ \emph {et~al.}(2013)\citenamefont {Long},
  \citenamefont {Wang}, \citenamefont {Ding}, \citenamefont {Sun},\ and\
  \citenamefont {Yu}}]{long2013transfer}%
  \BibitemOpen
  \bibfield  {author} {\bibinfo {author} {\bibfnamefont {M.}~\bibnamefont
  {Long}}, \bibinfo {author} {\bibfnamefont {J.}~\bibnamefont {Wang}}, \bibinfo
  {author} {\bibfnamefont {G.}~\bibnamefont {Ding}}, \bibinfo {author}
  {\bibfnamefont {J.}~\bibnamefont {Sun}},\ and\ \bibinfo {author}
  {\bibfnamefont {P.~S.}\ \bibnamefont {Yu}},\ }\bibfield  {title} {\bibinfo
  {title} {Transfer feature learning with joint distribution adaptation},\ }in\
  \href@noop {} {\emph {\bibinfo {booktitle} {Proceedings of the IEEE
  international conference on computer vision}}}\ (\bibinfo {year} {2013})\
  pp.\ \bibinfo {pages} {2200--2207}\BibitemShut {NoStop}%
\bibitem [{\citenamefont {Wang}\ \emph {et~al.}(2017)\citenamefont {Wang},
  \citenamefont {Chen}, \citenamefont {Hao}, \citenamefont {Feng},\ and\
  \citenamefont {Shen}}]{wang2017balanced}%
  \BibitemOpen
  \bibfield  {author} {\bibinfo {author} {\bibfnamefont {J.}~\bibnamefont
  {Wang}}, \bibinfo {author} {\bibfnamefont {Y.}~\bibnamefont {Chen}}, \bibinfo
  {author} {\bibfnamefont {S.}~\bibnamefont {Hao}}, \bibinfo {author}
  {\bibfnamefont {W.}~\bibnamefont {Feng}},\ and\ \bibinfo {author}
  {\bibfnamefont {Z.}~\bibnamefont {Shen}},\ }\bibfield  {title} {\bibinfo
  {title} {Balanced distribution adaptation for transfer learning},\ }in\
  \href@noop {} {\emph {\bibinfo {booktitle} {2017 IEEE International
  Conference on Data Mining (ICDM)}}}\ (\bibinfo {organization} {IEEE},\
  \bibinfo {year} {2017})\ pp.\ \bibinfo {pages} {1129--1134}\BibitemShut
  {NoStop}%
\bibitem [{\citenamefont {Blitzer}\ \emph {et~al.}(2006)\citenamefont
  {Blitzer}, \citenamefont {McDonald},\ and\ \citenamefont
  {Pereira}}]{blitzer2006domain}%
  \BibitemOpen
  \bibfield  {author} {\bibinfo {author} {\bibfnamefont {J.}~\bibnamefont
  {Blitzer}}, \bibinfo {author} {\bibfnamefont {R.}~\bibnamefont {McDonald}},\
  and\ \bibinfo {author} {\bibfnamefont {F.}~\bibnamefont {Pereira}},\
  }\bibfield  {title} {\bibinfo {title} {Domain adaptation with structural
  correspondence learning},\ }in\ \href@noop {} {\emph {\bibinfo {booktitle}
  {Proceedings of the 2006 conference on empirical methods in natural language
  processing}}}\ (\bibinfo {organization} {Association for Computational
  Linguistics},\ \bibinfo {year} {2006})\ pp.\ \bibinfo {pages}
  {120--128}\BibitemShut {NoStop}%
\bibitem [{\citenamefont {Blitzer}\ \emph {et~al.}(2007)\citenamefont
  {Blitzer}, \citenamefont {Dredze},\ and\ \citenamefont
  {Pereira}}]{blitzer2007biographies}%
  \BibitemOpen
  \bibfield  {author} {\bibinfo {author} {\bibfnamefont {J.}~\bibnamefont
  {Blitzer}}, \bibinfo {author} {\bibfnamefont {M.}~\bibnamefont {Dredze}},\
  and\ \bibinfo {author} {\bibfnamefont {F.}~\bibnamefont {Pereira}},\
  }\bibfield  {title} {\bibinfo {title} {Biographies, bollywood, boom-boxes and
  blenders: Domain adaptation for sentiment classification},\ }in\ \href@noop
  {} {\emph {\bibinfo {booktitle} {Proceedings of the 45th annual meeting of
  the association of computational linguistics}}}\ (\bibinfo {year} {2007})\
  pp.\ \bibinfo {pages} {440--447}\BibitemShut {NoStop}%
\bibitem [{\citenamefont {Fernando}\ \emph {et~al.}(2013)\citenamefont
  {Fernando}, \citenamefont {Habrard}, \citenamefont {Sebban},\ and\
  \citenamefont {Tuytelaars}}]{fernando2013unsupervised}%
  \BibitemOpen
  \bibfield  {author} {\bibinfo {author} {\bibfnamefont {B.}~\bibnamefont
  {Fernando}}, \bibinfo {author} {\bibfnamefont {A.}~\bibnamefont {Habrard}},
  \bibinfo {author} {\bibfnamefont {M.}~\bibnamefont {Sebban}},\ and\ \bibinfo
  {author} {\bibfnamefont {T.}~\bibnamefont {Tuytelaars}},\ }\bibfield  {title}
  {\bibinfo {title} {Unsupervised visual domain adaptation using subspace
  alignment},\ }in\ \href@noop {} {\emph {\bibinfo {booktitle} {Proceedings of
  the IEEE international conference on computer vision}}}\ (\bibinfo {year}
  {2013})\ pp.\ \bibinfo {pages} {2960--2967}\BibitemShut {NoStop}%
\bibitem [{\citenamefont {Gopalan}\ \emph {et~al.}(2011)\citenamefont
  {Gopalan}, \citenamefont {Li},\ and\ \citenamefont
  {Chellappa}}]{gopalan2011domain}%
  \BibitemOpen
  \bibfield  {author} {\bibinfo {author} {\bibfnamefont {R.}~\bibnamefont
  {Gopalan}}, \bibinfo {author} {\bibfnamefont {R.}~\bibnamefont {Li}},\ and\
  \bibinfo {author} {\bibfnamefont {R.}~\bibnamefont {Chellappa}},\ }\bibfield
  {title} {\bibinfo {title} {Domain adaptation for object recognition: An
  unsupervised approach},\ }in\ \href@noop {} {\emph {\bibinfo {booktitle}
  {2011 international conference on computer vision}}}\ (\bibinfo
  {organization} {IEEE},\ \bibinfo {year} {2011})\ pp.\ \bibinfo {pages}
  {999--1006}\BibitemShut {NoStop}%
\bibitem [{\citenamefont {Gong}\ \emph {et~al.}(2012)\citenamefont {Gong},
  \citenamefont {Shi}, \citenamefont {Sha},\ and\ \citenamefont
  {Grauman}}]{gong2012geodesic}%
  \BibitemOpen
  \bibfield  {author} {\bibinfo {author} {\bibfnamefont {B.}~\bibnamefont
  {Gong}}, \bibinfo {author} {\bibfnamefont {Y.}~\bibnamefont {Shi}}, \bibinfo
  {author} {\bibfnamefont {F.}~\bibnamefont {Sha}},\ and\ \bibinfo {author}
  {\bibfnamefont {K.}~\bibnamefont {Grauman}},\ }\bibfield  {title} {\bibinfo
  {title} {Geodesic flow kernel for unsupervised domain adaptation},\ }in\
  \href@noop {} {\emph {\bibinfo {booktitle} {2012 IEEE Conference on Computer
  Vision and Pattern Recognition}}}\ (\bibinfo {organization} {IEEE},\ \bibinfo
  {year} {2012})\ pp.\ \bibinfo {pages} {2066--2073}\BibitemShut {NoStop}%
\bibitem [{\citenamefont {Gretton}\ \emph {et~al.}(2007)\citenamefont
  {Gretton}, \citenamefont {Borgwardt}, \citenamefont {Rasch}, \citenamefont
  {Sch{\"o}lkopf},\ and\ \citenamefont {Smola}}]{gretton2007kernel}%
  \BibitemOpen
  \bibfield  {author} {\bibinfo {author} {\bibfnamefont {A.}~\bibnamefont
  {Gretton}}, \bibinfo {author} {\bibfnamefont {K.~M.}\ \bibnamefont
  {Borgwardt}}, \bibinfo {author} {\bibfnamefont {M.}~\bibnamefont {Rasch}},
  \bibinfo {author} {\bibfnamefont {B.}~\bibnamefont {Sch{\"o}lkopf}},\ and\
  \bibinfo {author} {\bibfnamefont {A.~J.}\ \bibnamefont {Smola}},\ }\bibfield
  {title} {\bibinfo {title} {A kernel method for the two-sample-problem},\ }in\
  \href@noop {} {\emph {\bibinfo {booktitle} {Advances in neural information
  processing systems}}}\ (\bibinfo {year} {2007})\ pp.\ \bibinfo {pages}
  {513--520}\BibitemShut {NoStop}%
\bibitem [{\citenamefont {Smola}\ \emph {et~al.}(2007)\citenamefont {Smola},
  \citenamefont {Gretton}, \citenamefont {Song},\ and\ \citenamefont
  {Sch{\"o}lkopf}}]{smola2007hilbert}%
  \BibitemOpen
  \bibfield  {author} {\bibinfo {author} {\bibfnamefont {A.}~\bibnamefont
  {Smola}}, \bibinfo {author} {\bibfnamefont {A.}~\bibnamefont {Gretton}},
  \bibinfo {author} {\bibfnamefont {L.}~\bibnamefont {Song}},\ and\ \bibinfo
  {author} {\bibfnamefont {B.}~\bibnamefont {Sch{\"o}lkopf}},\ }\bibfield
  {title} {\bibinfo {title} {A hilbert space embedding for distributions},\
  }in\ \href@noop {} {\emph {\bibinfo {booktitle} {International Conference on
  Algorithmic Learning Theory}}}\ (\bibinfo {organization} {Springer},\
  \bibinfo {year} {2007})\ pp.\ \bibinfo {pages} {13--31}\BibitemShut {NoStop}%
\bibitem [{\citenamefont {Pan}\ \emph {et~al.}(2008)\citenamefont {Pan},
  \citenamefont {Kwok}, \citenamefont {Yang} \emph {et~al.}}]{pan2008transfer}%
  \BibitemOpen
  \bibfield  {author} {\bibinfo {author} {\bibfnamefont {S.~J.}\ \bibnamefont
  {Pan}}, \bibinfo {author} {\bibfnamefont {J.~T.}\ \bibnamefont {Kwok}},
  \bibinfo {author} {\bibfnamefont {Q.}~\bibnamefont {Yang}}, \emph {et~al.},\
  }\bibfield  {title} {\bibinfo {title} {Transfer learning via dimensionality
  reduction.},\ }in\ \href@noop {} {\emph {\bibinfo {booktitle} {AAAI}}},\
  Vol.~\bibinfo {volume} {8}\ (\bibinfo {year} {2008})\ pp.\ \bibinfo {pages}
  {677--682}\BibitemShut {NoStop}%
\bibitem [{\citenamefont {Giovannetti}\ \emph {et~al.}(2008)\citenamefont
  {Giovannetti}, \citenamefont {Lloyd},\ and\ \citenamefont
  {Maccone}}]{giovannetti2008quantum}%
  \BibitemOpen
  \bibfield  {author} {\bibinfo {author} {\bibfnamefont {V.}~\bibnamefont
  {Giovannetti}}, \bibinfo {author} {\bibfnamefont {S.}~\bibnamefont {Lloyd}},\
  and\ \bibinfo {author} {\bibfnamefont {L.}~\bibnamefont {Maccone}},\
  }\bibfield  {title} {\bibinfo {title} {Quantum random access memory},\
  }\href@noop {} {\bibfield  {journal} {\bibinfo  {journal} {Physical review
  letters}\ }\textbf {\bibinfo {volume} {100}},\ \bibinfo {pages} {160501}
  (\bibinfo {year} {2008})}\BibitemShut {NoStop}%
\bibitem [{\citenamefont {Cong}\ and\ \citenamefont
  {Duan}(2016)}]{cong2016quantum}%
  \BibitemOpen
  \bibfield  {author} {\bibinfo {author} {\bibfnamefont {I.}~\bibnamefont
  {Cong}}\ and\ \bibinfo {author} {\bibfnamefont {L.}~\bibnamefont {Duan}},\
  }\bibfield  {title} {\bibinfo {title} {Quantum discriminant analysis for
  dimensionality reduction and classification},\ }\href@noop {} {\bibfield
  {journal} {\bibinfo  {journal} {New Journal of Physics}\ }\textbf {\bibinfo
  {volume} {18}},\ \bibinfo {pages} {073011} (\bibinfo {year}
  {2016})}\BibitemShut {NoStop}%
\bibitem [{\citenamefont {Duan}\ \emph {et~al.}(2017)\citenamefont {Duan},
  \citenamefont {Yuan}, \citenamefont {Liu},\ and\ \citenamefont
  {Li}}]{duan2017quantum}%
  \BibitemOpen
  \bibfield  {author} {\bibinfo {author} {\bibfnamefont {B.}~\bibnamefont
  {Duan}}, \bibinfo {author} {\bibfnamefont {J.}~\bibnamefont {Yuan}}, \bibinfo
  {author} {\bibfnamefont {Y.}~\bibnamefont {Liu}},\ and\ \bibinfo {author}
  {\bibfnamefont {D.}~\bibnamefont {Li}},\ }\bibfield  {title} {\bibinfo
  {title} {Quantum algorithm for support matrix machines},\ }\href@noop {}
  {\bibfield  {journal} {\bibinfo  {journal} {Physical Review A}\ }\textbf
  {\bibinfo {volume} {96}},\ \bibinfo {pages} {032301} (\bibinfo {year}
  {2017})}\BibitemShut {NoStop}%
\bibitem [{\citenamefont {Nielsen}\ and\ \citenamefont
  {Chuang}(2002)}]{nielsen2002quantum}%
  \BibitemOpen
  \bibfield  {author} {\bibinfo {author} {\bibfnamefont {M.~A.}\ \bibnamefont
  {Nielsen}}\ and\ \bibinfo {author} {\bibfnamefont {I.}~\bibnamefont
  {Chuang}},\ }\href@noop {} {\bibinfo {title} {Quantum computation and quantum
  information}} (\bibinfo {year} {2002})\BibitemShut {NoStop}%
\bibitem [{\citenamefont {Dervovic}\ \emph {et~al.}(2018)\citenamefont
  {Dervovic}, \citenamefont {Herbster}, \citenamefont {Mountney}, \citenamefont
  {Severini}, \citenamefont {Usher},\ and\ \citenamefont
  {Wossnig}}]{dervovic2018quantum}%
  \BibitemOpen
  \bibfield  {author} {\bibinfo {author} {\bibfnamefont {D.}~\bibnamefont
  {Dervovic}}, \bibinfo {author} {\bibfnamefont {M.}~\bibnamefont {Herbster}},
  \bibinfo {author} {\bibfnamefont {P.}~\bibnamefont {Mountney}}, \bibinfo
  {author} {\bibfnamefont {S.}~\bibnamefont {Severini}}, \bibinfo {author}
  {\bibfnamefont {N.}~\bibnamefont {Usher}},\ and\ \bibinfo {author}
  {\bibfnamefont {L.}~\bibnamefont {Wossnig}},\ }\bibfield  {title} {\bibinfo
  {title} {Quantum linear systems algorithms: a primer},\ }\href@noop {}
  {\bibfield  {journal} {\bibinfo  {journal} {arXiv preprint arXiv:1802.08227}\
  } (\bibinfo {year} {2018})}\BibitemShut {NoStop}%
\bibitem [{\citenamefont {Lloyd}(1996)}]{lloyd1996universal}%
  \BibitemOpen
  \bibfield  {author} {\bibinfo {author} {\bibfnamefont {S.}~\bibnamefont
  {Lloyd}},\ }\bibfield  {title} {\bibinfo {title} {Universal quantum
  simulators},\ }\href@noop {} {\bibfield  {journal} {\bibinfo  {journal}
  {Science}\ ,\ \bibinfo {pages} {1073}} (\bibinfo {year} {1996})}\BibitemShut
  {NoStop}%
\bibitem [{\citenamefont {Wang}\ \emph {et~al.}(2010)\citenamefont {Wang},
  \citenamefont {Wu}, \citenamefont {Liu},\ and\ \citenamefont
  {Nori}}]{wang2010measurement}%
  \BibitemOpen
  \bibfield  {author} {\bibinfo {author} {\bibfnamefont {H.}~\bibnamefont
  {Wang}}, \bibinfo {author} {\bibfnamefont {L.-A.}\ \bibnamefont {Wu}},
  \bibinfo {author} {\bibfnamefont {Y.-x.}\ \bibnamefont {Liu}},\ and\ \bibinfo
  {author} {\bibfnamefont {F.}~\bibnamefont {Nori}},\ }\bibfield  {title}
  {\bibinfo {title} {Measurement-based quantum phase estimation algorithm for
  finding eigenvalues of non-unitary matrices},\ }\href@noop {} {\bibfield
  {journal} {\bibinfo  {journal} {Physical Review A}\ }\textbf {\bibinfo
  {volume} {82}},\ \bibinfo {pages} {062303} (\bibinfo {year}
  {2010})}\BibitemShut {NoStop}%
\bibitem [{\citenamefont {Daskin}\ \emph {et~al.}(2014)\citenamefont {Daskin},
  \citenamefont {Grama},\ and\ \citenamefont {Kais}}]{daskin2014universal}%
  \BibitemOpen
  \bibfield  {author} {\bibinfo {author} {\bibfnamefont {A.}~\bibnamefont
  {Daskin}}, \bibinfo {author} {\bibfnamefont {A.}~\bibnamefont {Grama}},\ and\
  \bibinfo {author} {\bibfnamefont {S.}~\bibnamefont {Kais}},\ }\bibfield
  {title} {\bibinfo {title} {A universal quantum circuit scheme for finding
  complex eigenvalues},\ }\href@noop {} {\bibfield  {journal} {\bibinfo
  {journal} {Quantum information processing}\ }\textbf {\bibinfo {volume}
  {13}},\ \bibinfo {pages} {333} (\bibinfo {year} {2014})}\BibitemShut
  {NoStop}%
\bibitem [{\citenamefont {Du}\ \emph {et~al.}(2018{\natexlab{a}})\citenamefont
  {Du}, \citenamefont {Hsieh}, \citenamefont {Liu},\ and\ \citenamefont
  {Tao}}]{du2018expressive}%
  \BibitemOpen
  \bibfield  {author} {\bibinfo {author} {\bibfnamefont {Y.}~\bibnamefont
  {Du}}, \bibinfo {author} {\bibfnamefont {M.-H.}\ \bibnamefont {Hsieh}},
  \bibinfo {author} {\bibfnamefont {T.}~\bibnamefont {Liu}},\ and\ \bibinfo
  {author} {\bibfnamefont {D.}~\bibnamefont {Tao}},\ }\bibfield  {title}
  {\bibinfo {title} {The expressive power of parameterized quantum circuits},\
  }\href@noop {} {\bibfield  {journal} {\bibinfo  {journal} {arXiv preprint
  arXiv:1810.11922}\ } (\bibinfo {year} {2018}{\natexlab{a}})}\BibitemShut
  {NoStop}%
\bibitem [{\citenamefont {Roggero}\ and\ \citenamefont
  {Baroni}(2019)}]{roggero2019short}%
  \BibitemOpen
  \bibfield  {author} {\bibinfo {author} {\bibfnamefont {A.}~\bibnamefont
  {Roggero}}\ and\ \bibinfo {author} {\bibfnamefont {A.}~\bibnamefont
  {Baroni}},\ }\bibfield  {title} {\bibinfo {title} {Short-depth circuits for
  efficient expectation value estimation},\ }\href@noop {} {\bibfield
  {journal} {\bibinfo  {journal} {arXiv preprint arXiv:1905.08383}\ } (\bibinfo
  {year} {2019})}\BibitemShut {NoStop}%
\bibitem [{\citenamefont {Buhrman}\ \emph {et~al.}(2001)\citenamefont
  {Buhrman}, \citenamefont {Cleve}, \citenamefont {Watrous},\ and\
  \citenamefont {De~Wolf}}]{buhrman2001quantum}%
  \BibitemOpen
  \bibfield  {author} {\bibinfo {author} {\bibfnamefont {H.}~\bibnamefont
  {Buhrman}}, \bibinfo {author} {\bibfnamefont {R.}~\bibnamefont {Cleve}},
  \bibinfo {author} {\bibfnamefont {J.}~\bibnamefont {Watrous}},\ and\ \bibinfo
  {author} {\bibfnamefont {R.}~\bibnamefont {De~Wolf}},\ }\bibfield  {title}
  {\bibinfo {title} {Quantum fingerprinting},\ }\href@noop {} {\bibfield
  {journal} {\bibinfo  {journal} {Physical Review Letters}\ }\textbf {\bibinfo
  {volume} {87}},\ \bibinfo {pages} {167902} (\bibinfo {year}
  {2001})}\BibitemShut {NoStop}%
\bibitem [{\citenamefont {Garcia-Escartin}\ and\ \citenamefont
  {Chamorro-Posada}(2013)}]{garcia2013swap}%
  \BibitemOpen
  \bibfield  {author} {\bibinfo {author} {\bibfnamefont {J.~C.}\ \bibnamefont
  {Garcia-Escartin}}\ and\ \bibinfo {author} {\bibfnamefont {P.}~\bibnamefont
  {Chamorro-Posada}},\ }\bibfield  {title} {\bibinfo {title} {Swap test and
  hong-ou-mandel effect are equivalent},\ }\href@noop {} {\bibfield  {journal}
  {\bibinfo  {journal} {Physical Review A}\ }\textbf {\bibinfo {volume} {87}},\
  \bibinfo {pages} {052330} (\bibinfo {year} {2013})}\BibitemShut {NoStop}%
\bibitem [{\citenamefont {Cincio}\ \emph {et~al.}(2018)\citenamefont {Cincio},
  \citenamefont {Suba{\c{s}}{\i}}, \citenamefont {Sornborger},\ and\
  \citenamefont {Coles}}]{cincio2018learning}%
  \BibitemOpen
  \bibfield  {author} {\bibinfo {author} {\bibfnamefont {L.}~\bibnamefont
  {Cincio}}, \bibinfo {author} {\bibfnamefont {Y.}~\bibnamefont
  {Suba{\c{s}}{\i}}}, \bibinfo {author} {\bibfnamefont {A.~T.}\ \bibnamefont
  {Sornborger}},\ and\ \bibinfo {author} {\bibfnamefont {P.~J.}\ \bibnamefont
  {Coles}},\ }\bibfield  {title} {\bibinfo {title} {Learning the quantum
  algorithm for state overlap},\ }\href@noop {} {\bibfield  {journal} {\bibinfo
   {journal} {New Journal of Physics}\ }\textbf {\bibinfo {volume} {20}},\
  \bibinfo {pages} {113022} (\bibinfo {year} {2018})}\BibitemShut {NoStop}%
\bibitem [{\citenamefont {Zhang}\ \emph {et~al.}(2019)\citenamefont {Zhang},
  \citenamefont {Hsieh}, \citenamefont {Liu},\ and\ \citenamefont
  {Tao}}]{zhang2019quantum}%
  \BibitemOpen
  \bibfield  {author} {\bibinfo {author} {\bibfnamefont {K.}~\bibnamefont
  {Zhang}}, \bibinfo {author} {\bibfnamefont {M.-H.}\ \bibnamefont {Hsieh}},
  \bibinfo {author} {\bibfnamefont {L.}~\bibnamefont {Liu}},\ and\ \bibinfo
  {author} {\bibfnamefont {D.}~\bibnamefont {Tao}},\ }\bibfield  {title}
  {\bibinfo {title} {Quantum algorithm for finding the negative curvature
  direction in non-convex optimization},\ }\href@noop {} {\bibfield  {journal}
  {\bibinfo  {journal} {arXiv preprint arXiv:1909.07622}\ } (\bibinfo {year}
  {2019})}\BibitemShut {NoStop}%
\bibitem [{\citenamefont {Du}\ \emph {et~al.}(2018{\natexlab{b}})\citenamefont
  {Du}, \citenamefont {Hsieh}, \citenamefont {Liu},\ and\ \citenamefont
  {Tao}}]{du2018implementable}%
  \BibitemOpen
  \bibfield  {author} {\bibinfo {author} {\bibfnamefont {Y.}~\bibnamefont
  {Du}}, \bibinfo {author} {\bibfnamefont {M.-H.}\ \bibnamefont {Hsieh}},
  \bibinfo {author} {\bibfnamefont {T.}~\bibnamefont {Liu}},\ and\ \bibinfo
  {author} {\bibfnamefont {D.}~\bibnamefont {Tao}},\ }\bibfield  {title}
  {\bibinfo {title} {Implementable quantum classifier for nonlinear data},\
  }\href@noop {} {\bibfield  {journal} {\bibinfo  {journal} {arXiv preprint
  arXiv:1809.06056}\ } (\bibinfo {year} {2018}{\natexlab{b}})}\BibitemShut
  {NoStop}%
\bibitem [{\citenamefont {He}\ \emph {et~al.}(2019)\citenamefont {He},
  \citenamefont {Sun}, \citenamefont {Lyu},\ and\ \citenamefont
  {Wang}}]{he2019quantum}%
  \BibitemOpen
  \bibfield  {author} {\bibinfo {author} {\bibfnamefont {X.}~\bibnamefont
  {He}}, \bibinfo {author} {\bibfnamefont {L.}~\bibnamefont {Sun}}, \bibinfo
  {author} {\bibfnamefont {C.}~\bibnamefont {Lyu}},\ and\ \bibinfo {author}
  {\bibfnamefont {X.}~\bibnamefont {Wang}},\ }\bibfield  {title} {\bibinfo
  {title} {Quantum locally linear embedding},\ }\href@noop {} {\bibfield
  {journal} {\bibinfo  {journal} {arXiv preprint arXiv:1910.07854}\ } (\bibinfo
  {year} {2019})}\BibitemShut {NoStop}%
\bibitem [{\citenamefont {Cortes}\ and\ \citenamefont
  {Vapnik}(1995)}]{cortes1995support}%
  \BibitemOpen
  \bibfield  {author} {\bibinfo {author} {\bibfnamefont {C.}~\bibnamefont
  {Cortes}}\ and\ \bibinfo {author} {\bibfnamefont {V.}~\bibnamefont
  {Vapnik}},\ }\bibfield  {title} {\bibinfo {title} {Support-vector networks},\
  }\href@noop {} {\bibfield  {journal} {\bibinfo  {journal} {Machine learning}\
  }\textbf {\bibinfo {volume} {20}},\ \bibinfo {pages} {273} (\bibinfo {year}
  {1995})}\BibitemShut {NoStop}%
\bibitem [{\citenamefont {Nesterov}\ and\ \citenamefont
  {Nemirovskii}(1994)}]{nesterov1994interior}%
  \BibitemOpen
  \bibfield  {author} {\bibinfo {author} {\bibfnamefont {Y.}~\bibnamefont
  {Nesterov}}\ and\ \bibinfo {author} {\bibfnamefont {A.}~\bibnamefont
  {Nemirovskii}},\ }\href@noop {} {\emph {\bibinfo {title} {Interior-point
  polynomial algorithms in convex programming}}},\ Vol.~\bibinfo {volume} {13}\
  (\bibinfo  {publisher} {Siam},\ \bibinfo {year} {1994})\BibitemShut {NoStop}%
\end{thebibliography}%

\end{document}